\documentclass[prd,aps,twocolumn,showpacs,superscriptaddress]{revtex4-1}

\usepackage{array}
\usepackage{amsmath}
\usepackage{graphicx}
\usepackage{amssymb}

\usepackage{color}
\usepackage[colorlinks,citecolor=blue]{hyperref}
\usepackage{subfigure}
\usepackage{bm}
\usepackage[colorinlistoftodos]{todonotes}
\usepackage{mathtools}

\providecommand{\aap}{Astron.\ Astrophys.}
\providecommand{\apj}{Astrophys.\ J.}
\providecommand{\apjl}{Astrophys.\ J.}
\providecommand{\apjs}{Astrophys.\ J. Suppl.}
\providecommand{\physrep}{Phys.\ Rep.}
\providecommand{\pasa}{Publ.\ Astron.\ Soc.\ Austr.}

\providecommand{\jcap}{J.\ Cosmol.\ Astropart.\ Phys.}

\usepackage{fancyhdr}
\begin{document}

\title{Observable signatures of enhanced axion emission from protoneutron stars}

\author{Tobias~Fischer}\email{tobias.fischer@uwr.edu.pl}
\affiliation{Institute of Theoretical Physics, University of Wroc{\l}aw, 50-204 Wroc{\l}aw, Poland.}

\author{Pierluca~Carenza}\email{pierluca.carenza@ba.infn.it}
\affiliation{Dipartimento Interateneo di Fisica “Michelangelo Merlin”, Via Amendola 173, 70126 Bari, Italy.}
\affiliation{Istituto Nazionale di Fisica Nucleare--Sezione di Bari, Via Orabona 4, 70126 Bari, Italy.}

\author{Bryce~Fore}\email{bryce4@uw.edu}
\affiliation{Institute for Nuclear Theory, University of Washington, Seattle, Washington 98195, USA.}
\affiliation{Department of Physics, University of Washington, Seattle, Washington 98195, USA.}

\author{Maurizio~Giannotti}\email{mgiannotti@barry.edu}
\affiliation{Physical Sciences, Barry University, 11300 Northeast 2nd Avenue, Miami Shores, FL 33161, USA.
\\ \bigskip
}

\author{Alessandro~Mirizzi}\email{alessandro.mirizzi@ba.infn.it}
\affiliation{Dipartimento Interateneo di Fisica “Michelangelo Merlin”, Via Amendola 173, 70126 Bari, Italy.}
\affiliation{Istituto Nazionale di Fisica Nucleare--Sezione di Bari, Via Orabona 4, 70126 Bari, Italy.}

\author{Sanjay~Reddy}\email{sareddy@uw.edu}
\affiliation{Institute for Nuclear Theory, University of Washington, Seattle, Washington 98195, USA.}
\affiliation{Department of Physics, University of Washington, Seattle, Washington 98195, USA.}

\begin{abstract}
We perform general relativistic one-dimensional supernova (SN) simulations to identify observable signatures of enhanced axion emission from the pion-induced reaction $\pi^-+p\to n+a$ inside a newly born protoneutron star (PNS). We focus on the early evolution after the onset of the supernova explosion to predict the temporal and spectral features of the neutrino and axion emission during the first 10~s. Pions are included as explicit new degrees of freedom in hot and dense matter. Their thermal population and their role in axion production are both determined consistently to include effects due to their interactions with nucleons. For a wide range of ambient conditions encountered inside a PNS, we find that the pion-induced axion production dominates over nucleon-nucleon bremsstrahlung processes. By consistently including the role of pions on the dense matter equation of state and on the energy loss, our simulations predict robust discernible features of neutrino and axion emission from a galactic supernova that can be observed in terrestrial detectors. For axion couplings that are compatible with current bounds, we find a significant suppression with time of the neutrino luminosity during the first 10~s. This suggests that current bounds derived from the neutrino signal from SN 1987A can be improved, and that future galactic supernovae may provide significantly more stringent constraints. 
\\
\smallskip
\\
DOI: 10.1103/PhysRevD.104.103012
\\
\end{abstract}

\received{12 August 2021}\accepted{11 October 2021}\published{9 November 2021}

\maketitle

\section{Introduction}
\label{sec:intro}
Core-collapse supernovae (SN) are ideal high-energy nuclear and particle physics laboratories, with baryon densities reaching a few times nuclear saturation density at their interiors and temperatures of several tens of MeV, as well as large isospin asymmetry~\cite{Fischer:2017}. The latter is quantified by the proton abundance ranging from extremely neutron-rich conditions with $Y_p\simeq 0.05$ to isospin symmetric conditions with $Y_p=0.5$. The central compact stellar object of a core-collapse SN is the protoneutron star (PNS). It forms when the collapsing stellar core of a massive star bounces back at densities slightly in excess of nuclear saturation density ($\rho_{\rm sat}\simeq 2.5\times 10^{14}$~g~cm$^{-3}$). The later postbounce evolution of the PNS is determined by continuous mass accretion, from the still-collapsing outer layers of the progenitor star. When the SN explosion develops---because of the revival of the stalled bounce shock---the nascent PNS will deleptonize via the emission of neutrinos of all flavors on a timescale of several tens of seconds~\cite{Pons:1998mm} and later cool to form the final SN remnant neutron star. While for the understanding of the SN explosion, in particular aided through the neutrino-driven paradigm, multidimensional phenomena play a crucial role~\cite{Janka:2007,Mirizzi:2016}, the PNS deleptonization is well approximated within spherically symmetric simulation studies~\cite{Huedepohl:2010,Fischer:2009af,Roberts:2012b,Roberts:2012c,Fischer:2016a}. 

With the recent advances in the modeling of core-collapse SN, which now allow the inclusion of muons as the degree of freedom and to treat six-species Boltzmann neutrino transport schemes~\cite{Bollig:2017,Fischer:2020d}, it becomes possible to study novel high-energy astroparticle physics phenomena. Besides muons the next heavier degrees of freedom to consider are the pions. These bosons exist as charge-neutral $\pi^0$ and as charged $\pi^\pm$. Pions are one of the key messengers in the context of heavy-ion collisions (for a recent comprehensive review, see Ref.~\cite{Redlich:2018}), which makes them one of the best-studied particle degrees of freedom. The role of pions in neutron stars has been studied for a long time, particularly in relation to the emergence of a pion condensate. However, in the context of core-collapse SN, only the core-collapse SN simulations of Ref.~\cite{Wilson:1993rx} included pions in the equation of state (EOS). Recently, the {\em classical} neutron star EOS of Akmal, Pandharipande and Ravenhall, denoted as APR~\cite{AP:1997,APR:1998}, was extended to finite temperatures and arbitrary isospin asymmetry applying a Skyrme functional model, prepared for astrophysical applications~\cite{Schneider:2019}.  Pions are included through the phase transition to a neutral pion condensate at supersaturation density.  This results in a softening of the EOS. As was pointed out in Ref.~\cite{Fore:2020}, an even greater impact on the high-density EOS emerges already below saturation density when considering the contributions from strongly interacting pions, giving rise to the pion self-energy. The resulting pion yields can be comparable to those of muons, depending on the nuclear EOS. In this article we employ this formalism, based on the virial EOS featuring pion-nucleon scattering phase shifts, and study the appearance of pions in simulations of the PNS deleptonization.

The inclusion of pions in the EOS allows us to describe, for the first time, the full impact of axions on the SN evolution in a self-consistent way. Axions are a prediction of the most promising solution of the {\em strong CP problem}, which is related to the absence of the expected {\em CP} violation in the strong interactions~\cite{Peccei:1977hh,Peccei:1977ur,Weinberg:1977ma,Wilczek:1977pj}. The axion phenomenology is widely discussed in the literature (for recent reviews, see Refs.~\cite{DiVecchia:2019ejf,DiLuzio:2020wdo,Agrawal:2021dbo}). It is essential to remark that axions are expected to couple to photons, electrons, and nucleons, with model-dependent couplings~\cite{diCortona:2015ldu,Borsanyi:2016ksw}. These can be constrained through laboratory experiments~\cite{Irastorza:2018dyq,Sikivie:2020zpn} and astrophysical considerations~\cite{Raffelt:2006cw,Giannotti:2015kwo,Giannotti:2017hny}.
In particular, core-collapse SN have long been used to constrain the axion couplings to nucleons~\cite{Carenza:2019pxu},  photons~\cite{Payez:2014xsa,Calore:2020tjw}, and, recently, to muons~\cite{Bollig:2020}. In the present paper, we are particularly interested in axions coupled to nucleons. In this case, the most widely studied mechanism for axion production in a SN core is the nucleon-nucleon bremsstrahlung, 
$N+N\to N+N+a$~\cite{Raffelt:1987yt,Turner:1987by,Burrows:1988ah,Burrows:1990pk,Raffelt:1990yz,Carenza:2019pxu}. 
Because of the efficient energy loss through axions, the above process contributes to the shortening of the neutrino emission from a core-collapse SN. Quantitatively, this time reduction is parameterized by axion-nucleon coupling~\cite{Fischer:2016b}. The current bound on the axion-nucleon coupling has been revisited~\cite{Carenza:2019pxu}, with improved nucleon-nucleon bremsstrahlung rates for the axion emissivity taking into account leading-order medium modifications. Furthermore, in the context of core-collapse SN, there is a second competing process for the emission of axions, which has long been omitted, namely the emission of axions from Compton pionic processes, $\pi^-+p\to n+a$~\cite{Turner:1988,Raffelt:1993ix,Keil:1996ab}. This process has been reviewed recently in the context of strongly interacting pions under the conditions of hot neutron stars~\cite{Carenza:2020}. It leads to the conclusion that the previous assumption of the dominating axion emission from nucleon-nucleon bremsstrahlung over those from pions must be relaxed. 

In order to quantify the impact of axion losses originating from pions, the present study implements the associated axion emissivities~\cite{Carenza:2020} in simulations of the PNS deleptonization phase of a core-collapse SN explosion, together with the updated axion emissivities from nucleon-nucleon bremsstrahlung~\cite{Carenza:2019pxu}. Particular emphasis is devoted to the impact of the shortening of the neutrino emission due to the associated enhanced cooling, and the potentially observable signature at the future-planned Hyper-Kamiokande and presently operating Super-Kamiokande water-Cherenkov detectors.

The paper is organized as follows.  In Sec.~\ref{sec:SNmodel}, we review our core-collapse SN model  with the necessary updates to include pions. In Sec.~\ref{sec:rates}, we introduce the axion emission processes, specifically the improved nucleon-nucleon bremsstrahlung rates and the pion rates. Their impact on the PNS deleptonization will be discussed in Sec.~\ref{sec:PNS_delept_axions}. In Sec.~\ref{sec:detection} we present the observable signatures of the axion emission and discuss the impact on the neutrino signal. The paper closes with a summary in Sec.~\ref{sec:summary}.

\section{Supernova model with pions}
\label{sec:SNmodel}
The spherically symmetric core-collapse SN model employed in this study, {\tt AGILE-BOLTZTRAN}, is based on general relativistic neutrino radiation hydrodynamics~\cite{Mezzacappa:1993gm,Mezzacappa:1993gn,Mezzacappa:1993gx,Liebendoerfer:2004}. It includes an adaptive baryon mass mesh refinement~\cite{Liebendoerfer:2002,Fischer:2009af}, where for the present study we use 207 radial mass shell grid points. {\tt AGILE-BOLTZTRAN} has been extended to handle six-species Boltzmann neutrino transport~\cite{Fischer:2020d} with the inclusion of the muon abundance $Y_\mu$, as an additional degree of freedom, together with a comprehensive set of muonic weak interactions in the collision integral of the Boltzmann transport equation~\cite{Guo:2020,Fischer:2020d}. The list of standard, nonmuonic weak processes used here can be found in Table~I of Ref.~\cite{Fischer:2020a}, with the charged-current weak rates, including the (inverse) neutron decay channel, in the full-kinematics treatment as well as including self-consistently contributions from weak magnetism. The latter two are particularly important for the PNS deleptonization phase, as was discussed in Ref.~\cite{Fischer:2020a}. The neutrino distributions and transport are discretized with six momentum angles, $\cos\vartheta\in[-1,1]$, and 36 neutrino energy bins, $E_\nu\in[0.5,300]$~MeV, following the setup of Bruenn~\cite{Bruenn:1985en}.

{\tt AGILE-BOLTZTRAN} has a flexible EOS module which can handle a variety of different nuclear matter EOS~\cite{Lattimer:1991nc,Shen:1998gg,Hempel:2009mc,Hempel:2011,Steiner:2013}. In the present work the DD2 relativistic mean-field EOS with density-dependent nucleon-meson coupling is employed \cite{Typel:2005,Typel:2009sy,Typel:2013rza}, together with the modified nuclear statistical equilibrium EOS of Ref.~\cite{Hempel:2009mc} for the description of medium nuclear clusters, henceforth denoted as HS(DD2). It provides a good description of nuclear saturation properties as well as neutron star configurations, consistent with current maximum mass constraints from the observations of massive pulsars of about 2~$M_\odot$~\cite{Antoniadis:2013,Fonseca:2016}. Furthermore, electrons, positrons and photons are treated following Ref.~\cite{Timmes:1999}, while the muon EOS is based on a tabulation~\cite{Fischer:2020d}. 

The present work extends previous studies of axion emission in core-collapse SN~\cite{Fischer:2016b}, which exclusively focused on nucleon-nucleon bremsstrahlung. Here, we add a novel process of axion emission involving pions~\cite{Carenza:2020}.  To study self-consistently the effects of this process in the SN evolution, pions have to be included in the SN simulation. Pions have a mass of $m_{\pi^{0,\pm}}=135(140)$~MeV and are, hence, only slightly more massive than muons. Unlike muons, which are not in chemical equilibrium and, hence. are exclusively produced from weak processes under SN conditions, pions are in chemical equilibrium through strong interactions. Hence, it is not required to solve an evolution equation for the pion abundances, as is the case with muons~\cite{Bollig:2017,Fischer:2020d}. Instead, their abundances can be calculated from the corresponding local thermal and chemical equilibrium conditions for a gas of massive and relativistic bosons. This implies that the pions have a chemical potential, $\mu_{\pi^\pm}=\mp \hat\mu$, where $\hat\mu\equiv\mu_n-\mu_p$ is the chemical potential for negative charge and $\mu_{\pi^0}=0$.

\begin{figure}[t!]
\centering
\vspace{5mm}
\includegraphics[width=\columnwidth]{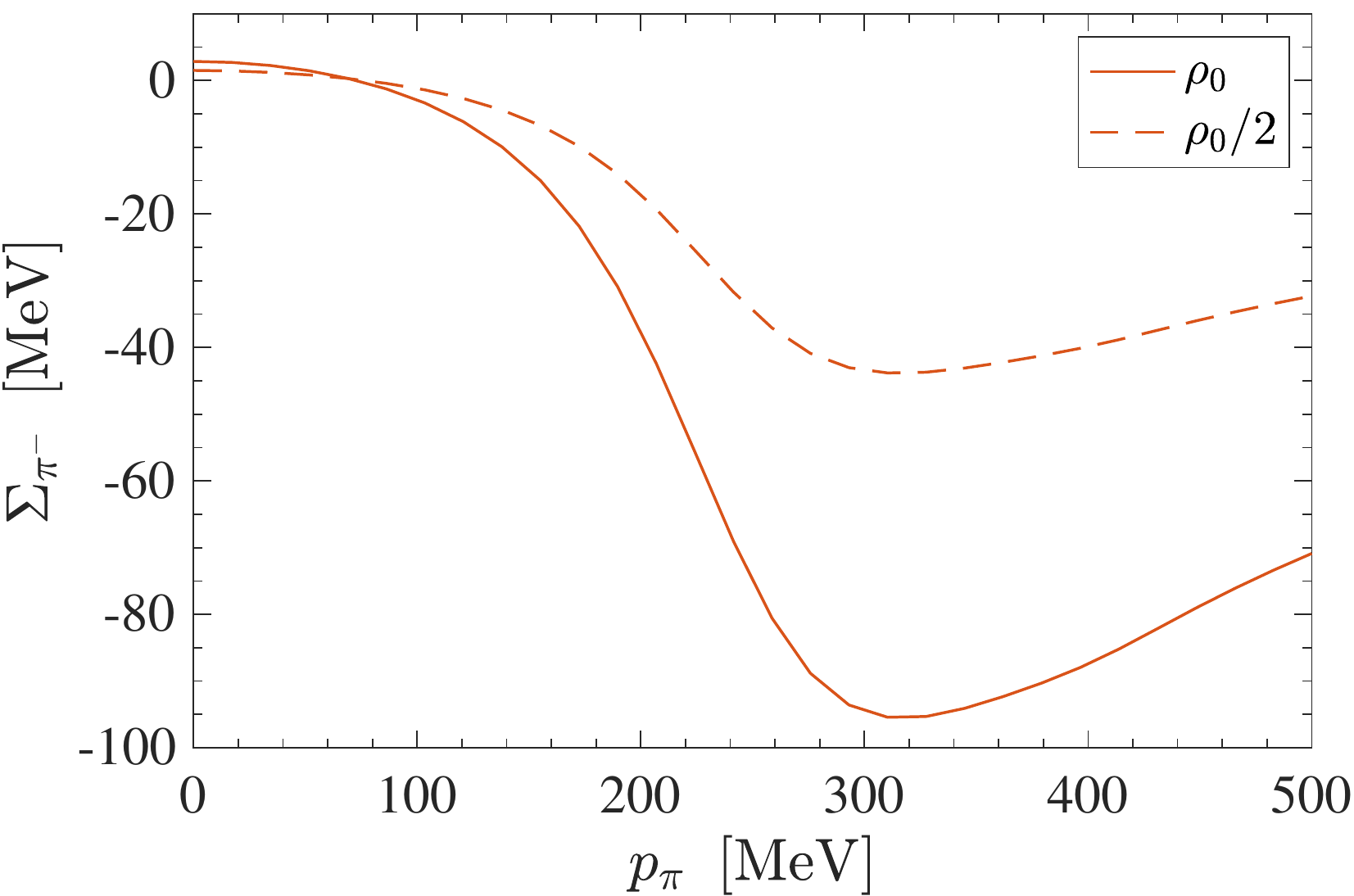}
\caption{Momentum $p_\pi$-dependent pion self-energy for $\pi^-$, $\Sigma_{\pi^-}$, calculated based on the nucleon-nucleon potential, Eqs.~(21)--(22) in Ref.~\cite{Fore:2020}, evaluated at a temperature of $T=20$~MeV and at fixed electron and muon abundances of $Y_e=0.15$ and $Y_\mu=0.04$, as well as two different baryon densities, at saturation density (solid line) and at one-half of saturation density (dashed line).}
\label{fig:sigma}
\end{figure}
\begin{figure}[t!]
\centering
\includegraphics[width=\columnwidth]{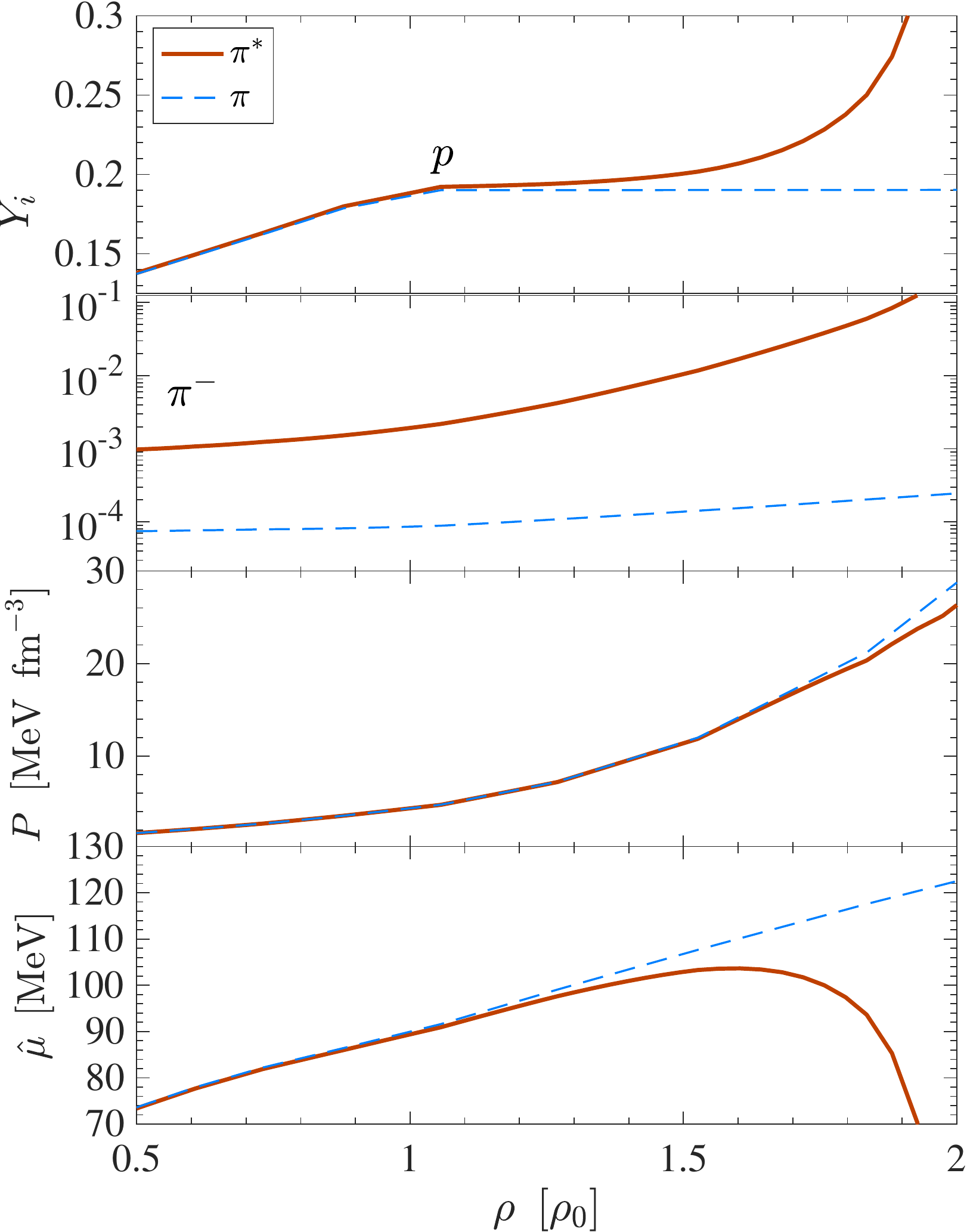}
\caption{EOS comparison, showing the particle yields $Y_i$ of protons ($p$) and pions ($\pi^-$), the total pressure $P$, and the charged chemical potential $\hat\mu$, as a function of the baryon density $\rho$, in units of the saturation density $\rho_0$, comparing the model with only thermal pions (blue dashed lines) and with interacting pions (brown solid lines), for the  nucleon-nucleon potential \citep[Eqs.~(21) and (22) in Ref.][]{Fore:2020}, evaluated at a temperature of $T=20$~MeV and fixed electron and muon abundances of $Y_e=0.15$ and $Y_\mu=0.04$.}
\label{fig:eos}
\end{figure}

Furthermore, pions are subject to strong interactions. At low density and high temperature, the virial expansion provides an efficient and model-independent approach to include pion-nucleon interactions. Under these conditions, when the pion and nucleon fugacities are small, the modification of the pion density and thermodynamics due to their interactions with neutrons and protons is adequately incorporated through the second virial coefficients, $b_2^{n\pi}$ and $b_2^{p\pi}$, respectively. We employ the methods described in Ref.~\cite{Fore:2020} to calculate these coefficients using measured pion-nucleon scattering phase shifts. 

To describe reactions involving pions and nucleons, we will need in addition a model for the pion dispersion relation in the hot and dense medium. We adopt the simple pseudopotential model described in Ref.~\cite{Fore:2020} and write the pion energy as 
\begin{equation}
E_{\pi^-} = \sqrt{p^2 + m_{\pi^-}^2} + \Sigma_{\pi^-}(p_\pi)~\,,
\label{eq:pion-selfenergy}
\end{equation}
where $\Sigma_{\pi^-}(p_\pi),$ is the momentum $p_pi$-dependent real part of the pion self-energy. The strength of the pseudopotential is tuned to ensure that the number densities obtained are consistent with the virial expansion\cite{Fore:2020}. 
The model includes the weakly repulsive s-wave interaction and the strongly attractive p-wave interactions between pions and nucleons. 
The latter has the effect of enhancing the $\pi^-$ and proton abundances (see Fig.~2 in Ref.~\cite{Fore:2020} and make a substantial contribution to the EOS. Since the pion self-energy $\Sigma_{\pi^-}$ uses the knowledge of the nucleon's equilibrium Fermi-Dirac distribution functions, with the associated nucleon chemical potentials, together with the pion-nucleon potential---see expressions~(21) and (22) in Ref.~\cite{Fore:2020}---it becomes evident that interaction contributions have to be consistent with the underlying nuclear HS(DD2) EOS. The pion self-energies are shown in Fig.~\ref{fig:sigma} for selected conditions relevant for the PNS deleptonization. It becomes clear that their contributions are indeed significant and cannot be neglected, neither for the calculation of the pion contributions to the EOS nor for the emission of axions stemming from pions, as will be discussed further below.

Note that the yields of $\pi^+$ are strongly suppressed due to the contributions of the Boltzmann suppression term,  $Y_{\pi^+}\propto\exp\{-\hat\mu/T\}$, in comparison to $Y_{\pi^-}\propto\exp\{\hat\mu/T\}$. A similar suppression is found for $\pi^0$, in comparison to $\pi^-$. Furthermore, we have  calculated the self-energies for $\pi^+$ which are found to be similar to those of $\pi^-$, and, hence, we find that at all conditions, the abundances of $\pi^+$ and $\pi^0$ are indeed negligible.

With the inclusion of pions, the charge neutrality condition is modified. It results in the following definition of the proton abundance: $Y_p = Y_e + Y_\mu + Y_{\pi^-} - Y_{\pi^+}$, which is one of the three independent nuclear EOS variables, in addition to temperature and baryon (or rest mass) density.  This results in a feedback to the baryon EOS, which is illustrated in Fig.~\ref{fig:eos} for a selected temperature of $T=20$~MeV, an electron fraction and a muon fraction of $Y_e=0.15$ and $Y_\mu=0.04$, lowering the abundance of neutrons and increasing the abundance of protons. For the calculation of the pion pressure and energy density, we used expressions~(9) and (10) of Ref.~\cite{Fore:2020}.

The question about weak interactions involving pions, i.e. the $\pi^-$ decay and its inverse, $\bar\nu_\mu + \mu^- \leftrightarrows \pi^-$ and $\nu_\mu+\pi^- \leftrightarrows \mu^-$, has already been addressed in Ref.~\cite{Fore:2020}. Note that the latter is possible only in the presence of interacting pions, i.e., when $\Sigma_{\pi^-}$ is large and negative to ensure energy and momentum conservation.  These reactions are competing not only with the neutral current scattering processes, as was discussed in Ref.~\cite{Fore:2020}, but also with the muonic charged-current reactions as well as the purely leptonic reaction channels~\cite{Guo:2020,Fischer:2020d}, giving additional contributions to the $\nu_\mu$ and $\bar\nu_\mu$ opacity accordingly.

The conditions for these weak processes requires a finite abundance of $\pi^-$. This is the case  at densities in excess of about $\rho\simeq10^{13}$~g~cm$^{-3}$. This, in turn, corresponds to temperatures fairly above $T>10$~MeV during the SN evolution, where the muon (anti)neutrinos are trapped entirely~\cite{Fischer:2012a}. As was found in Ref.~\cite{Fore:2020}, these weak processes would contribute mostly to low muon-(anti)neutrino energies. Hence, it may leave an imprint on the long-term muon-neutrino signal during the later PNS cooling phase, i.e. when the temperature drops at the PNS surface and the neutrinospheres shift to higher densities continuously. Here, for the study of the early PNS deleptonization phase for no more than about 10~s, the inclusion of weak reactions with pions has a negligible impact on the neutrino emission. 

Results from SN simulations which include muons and pions, the latter both noninteracting and interacting, are presented in the Appendix~\ref{sec:PNS_delept_ref}, investigating the impact of muons and pions on the early PNS deleptonization phase. In comparison with the reference case without muons and pions, no significant deviation could be found for the models including muons and pions for the first 20~s of the PNS deleptonization, independent of the treatment of pions. 

The implementation of the axion losses in {\tt AGILE-BOLTZTRAN} as a sink term in the evolution equation for the internal energy, corresponding to the conservation of energy, is discussed in Ref.~\cite{Fischer:2016b}. Here, the same approach is employed and we add to the previously included axion emission from nucleon-nucleon bremsstrahlung the updated rates as well as those from pions. Freely streaming axions are assumed. The axion-emission rates will be introduced and discussed in the next section.

\begin{figure*}[htp]
\centering
\includegraphics[width=1.9\columnwidth]{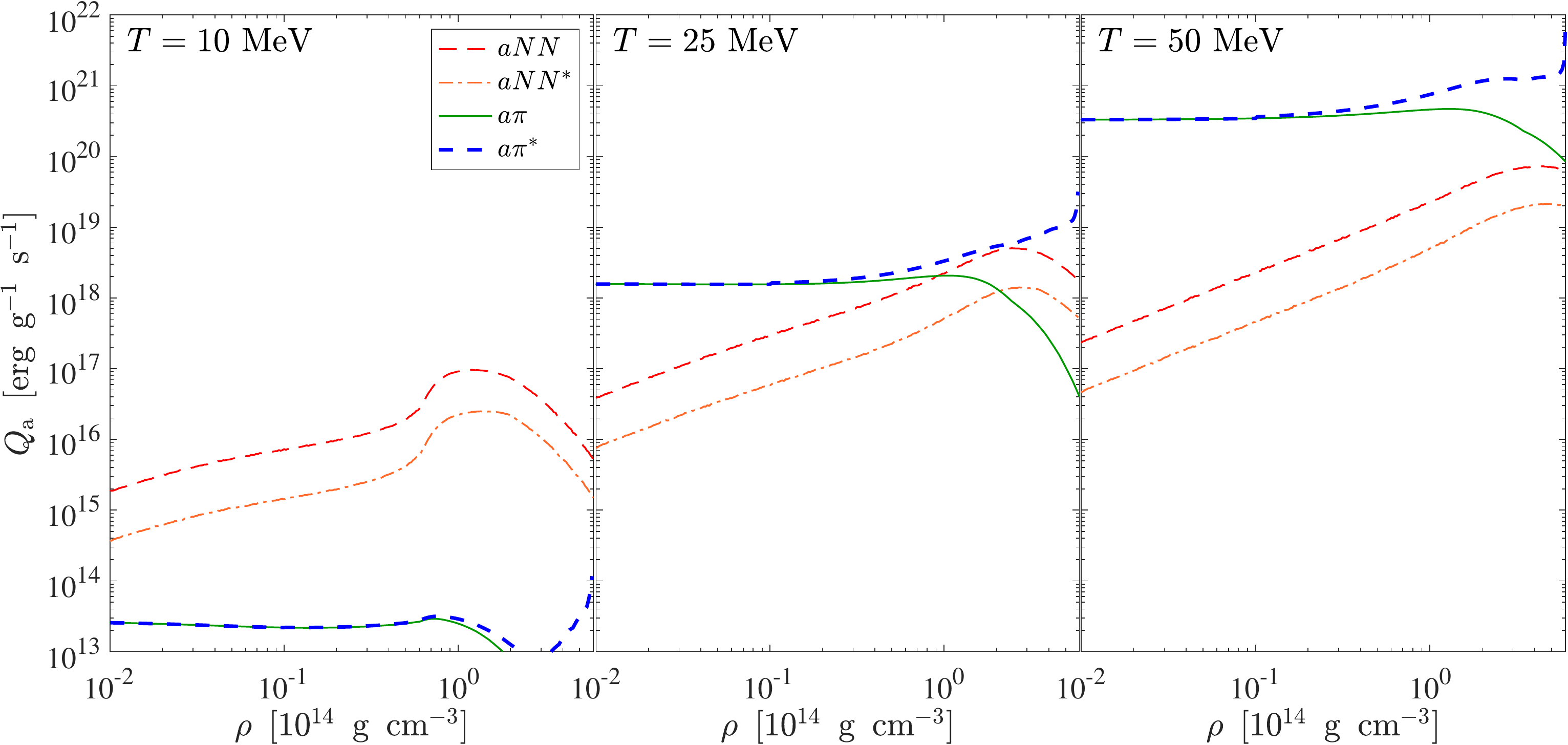}
\caption{Axion emissivities $Q_{\rm a}$, with respect to the rest mass density for fixed electron and muon abundances of $Y_e=0.3$ and $Y_\mu=10^{-4}$, respectively, as well as for three selected temperatures of $T=10$ (left panel), 25 (middle panel), and 50~MeV (right panel), comparing the bremsstrahlung rates $aNN$ (thin red dashed lines) and $aNN^*$ (thin orange dash-dotted lines) as well as axions stemming from pions $a\pi$ (thin green solid lines) and $a\pi^*$ (thick blue dashed lines).}
\label{fig:axion_rates}
\end{figure*}
%

\section{Axion-emission rates}
\label{sec:rates}
At present, the process for axion emission in the context of core-collapse SN has long been the nucleon-nucleon bremsstrahlung,
\begin{equation}
N+N\longrightarrow N+N+a~.
\label{eq:NNa}
\end{equation}
The associated rate expression for the axion emissivity is a complicated high-dimensional momentum integral, involving the matrix elements with the yet incompletely known axion-nucleon coupling constant $g_{aN}$ \citep[c.f. expressions~(2.2) and (2.6) in Ref.][and references therein]{Carenza:2019pxu}. Previously employed simplifications, such as the vacuum one-pion exchange  potential to describe the nuclear interactions and the assumption of massless pions, have lead to a simplified and semianalytical expression for the axion emissivity~\cite{Brinkmann:1988vi,Raffelt:1993ix}, henceforth denoted as $aNN$. This treatment was previously implemented in simulations of core-collapse SN~\cite{Fischer:2016b}. It was found that $aNN$ leads to an overestimation of the axion emission rate~\cite{Carenza:2019pxu}, which, hence, artificially enhances the axion losses in simulations of core-collapse SN. This caveat has been overcome with updated axion-emission rates derived in Ref.~\cite{Carenza:2019pxu}, henceforth denoted as $aNN^*$. It includes a substantially improved description of the nuclear interactions, i.e., nonvanishing mass for the exchange pions, contributions from the two-pion exchange, effective in-medium nucleon masses, and multiple nucleon scattering effects, as well as the proper phase-space contributions including final-state Pauli blocking terms. The latter are realized numerically via a 64-point Gauss-quadrature integration. Note that the same nuclear model has been the foundation for the updated neutrino-pair emission and absorption rates from $N$--$N$ bremsstrahlung~\cite{Guo:2019}. Globally, they provide a substantial reduction of the axion emissivity with increasing density and temperature. The present paper implements the $aNN^*$ treatment into the SN model {\tt AGILE-BOLTZTRAN}. Together with the previously employed simplistic axion emissivity, $aNN$, this will allow us to quantify the impact of the medium modifications from the associated axion cooling contributions.

Another very efficient axion production mechanism is the pion Compton scattering
\begin{equation}
\pi^- + p\longrightarrow n+a~.
\label{eq:pia}
\end{equation}
Reactions involving $\pi^+$ and $\pi^0$ are also possible but they are strongly suppressed compared to $\pi^-$ since the abundances of $\pi^+$ and $\pi^0$ are negligible, as was discussed in Sec.~\ref{sec:SNmodel}. The caveat of previous studies of axion emission from pions was the generally too low pion abundance, also for $\pi^-$, to be compatible with axions from nucleon-nucleon bremsstrahlung~\cite{Keil:1996ab}. However, the medium-modified suppression of the bremsstrahlung rates leads to the revision of this argument~\cite{Carenza:2020}. The subsequent microscopic axion-emission rate allows us to further support this argument, taking into account the treatment of interacting pions~\cite{Fore:2020}. It results not only in the substantial enhancement of the pion abundances at high density but also in an axion emissivity which is compatible to those of the bremsstrahlung processes. The axion emissivities of Ref.~\cite{Carenza:2020} are included into {\tt AGILE-BOLTZTRAN}, complementary to the bremsstrahlung rates. 

Important here is the treatment of the interacting contributions to the pions consistently with the underlying HS(DD2) EOS used in the simulations. Therefore, the computation of the axion emissivity for the process \eqref{eq:pia} implements the pion dispersion relation~\eqref{eq:pion-selfenergy}, including the momentum dependent pion self-energies [see Eq.~\eqref{eq:pion-selfenergy} and Fig.~\ref{fig:sigma}]. The axion emissivity through pion processes is henceforth denoted as $a\pi$, in the case of vanishing pion interactions, where we set $\Sigma_{\pi^-}=0$, and as $a\pi^*$ in the general case of interacting pions.

\begin{table*}[t!]
\caption{Summary of the different supernova simulations including the references to the various treatments for the calculation of the axion emissivity.}
\centering
\begin{tabular}{l cccc c cccc c}
\hline
\hline
Label & &&&& $N + N\rightarrow N + N + a$ & &&&& $\pi^- + p \rightarrow n + a$ \\
\hline
Ref. run (Appendix~\ref{sec:PNS_delept_ref}) &&&&& \ldots &&&&& \ldots \\
$aNN$ &&&&& Vacuum one-$\pi$ exchange, $m_\pi=0$ \cite{Brinkmann:1988vi,Raffelt:1993ix,Fischer:2016b} &&&&& \ldots \\
$aNN^*$ &&&&& Improvements according to Ref.~\cite{Carenza:2019pxu} &&&&& \ldots \\
$aNN^* + a\pi$ &&&&& Improvements according to Ref.~\cite{Carenza:2019pxu}  &&&&& Rates according to Ref.~\cite{Carenza:2020} with $\Sigma_\pi = 0$ \\
$aNN^* + a\pi^*$ &&&&& improvements according to Ref.~\cite{Carenza:2019pxu}  &&&&& Rates according to Ref.~\cite{Carenza:2020}, \\
&&&&&  &&&&& with $\Sigma_\pi$ according to Ref.~\cite{Fore:2020} \\
\hline
\end{tabular}
\label{tab:SN_setup}
\end{table*}

Figure~\ref{fig:axion_rates} compares the axion emissivity, denoted as $Q_{\rm a}$, which is the rate divided by the rest mass density and, hence, reflects the asymptotic behavior at low density which is proportional to the number density of targets, i.e., the abundance of $\pi^-$ for the axions stemming from process~\eqref{eq:pia} and $NN$ pairs for the bremsstrahlung processes \eqref{eq:NNa}. The latter is similar as the low-density behavior of the neutrino pair production from nucleon-nucleon bremsstrahlung processes (see Fig.~3 in Ref.~\cite{Fischer:2016a}). Guided by the previously revisited constraints for the axion-nucleon coupling constant~\cite{Carenza:2019pxu}, we choose here $g_{ap}=1.2\times 10^{-9}$ and $g_{an}=0$ for the calculations of the axion emissivities for the reactions \eqref{eq:NNa} and \eqref{eq:pia}, since the same axion-nucleon coupling gives rise to the axion emission for the bremsstrahlung processes and the pion conversion. We will employ the same values for the coupling constants in the SN simulations, which will be discussed below in Sec.~\ref{sec:PNS_delept_axions}. Here, we compare the different treatments $aNN$ and $aNN^*$ for several temperatures, $T=10$~(left panel), 25~(middle panel), and 50~MeV~(right panel). From this analysis becomes evident not only the strong temperature dependence of this axion-emission process but also the substantial overestimation of the axion emissivity for the simplified treatment $aNN$. It is important to note that the different axion emissivities depend on the nuclear EOS, for which we use for the HS(DD2) EOS for all calculations and SN simulations.

Figure~\ref{fig:axion_rates} compares the bremsstrahlung rates with the axion emissivity from Compton pionic processes, for $a\pi$ (thin solid green lines) and $a\pi^*$ (thick blue dashed lines). There are several observations: {\rm (i)} At low temperatures, there is no significant axion emission, neither from bremsstrahlung nor stemming from pions; {\rm (ii)} the overall low-density behavior reflects the limiting dependence on the temperature of the pion abundance; {\rm (iii)} the match between $a\pi$ and $a\pi^*$ at low density as strongly interacting pions divert into noninteracting pions; and {\rm (iv)} the substantial enhancement of the emissivity for $a\pi^*$ in comparison to $a\pi$ at a density in excess of $\rho\simeq10^{14}$~g~cm$^{-3}$, dominating the axion emission from pions over those of bremsstrahlung, however, only at high temperatures. 

\begin{figure*}[t!]
\centering
\includegraphics[width=1.35\columnwidth]{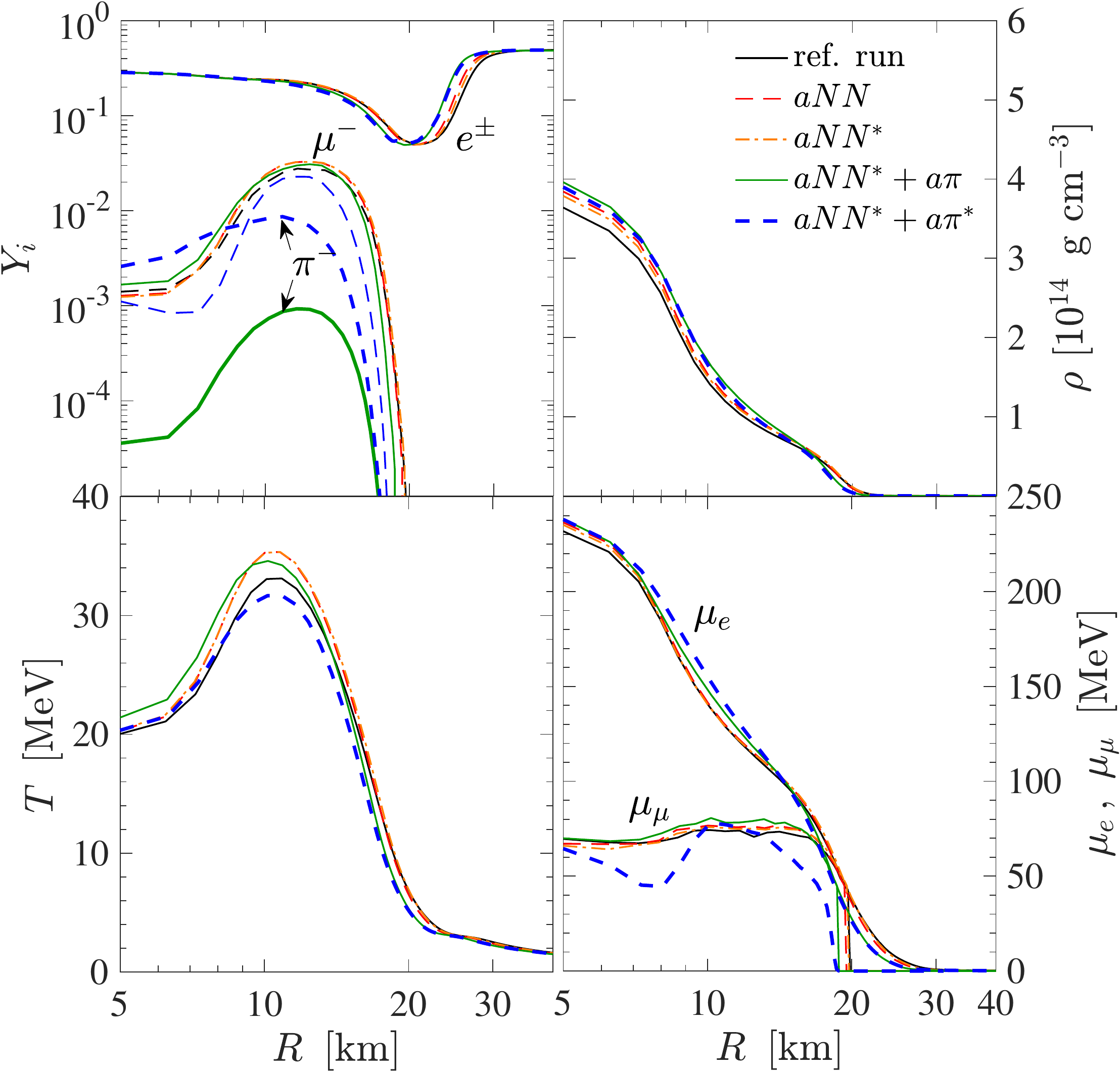}
\caption{Radial profiles of selected quantities, fractions of electrons ($Y_e$),  muons ($Y_\mu$), and pions ($Y_{\pi^-}$), temperature ($T$), rest mass density ($\rho$) and the chemical potentials of electrons ($\mu_e$) and muons ($\mu_\mu$) as well as the nuclear charge chemical potential ($\hat\mu$) at about 1~s postbounce during the early PNS deleptonization phase.}
\label{fig:hydro_a}
\end{figure*}
\begin{figure*}[t!]
\centering
\includegraphics[width=1.35\columnwidth]{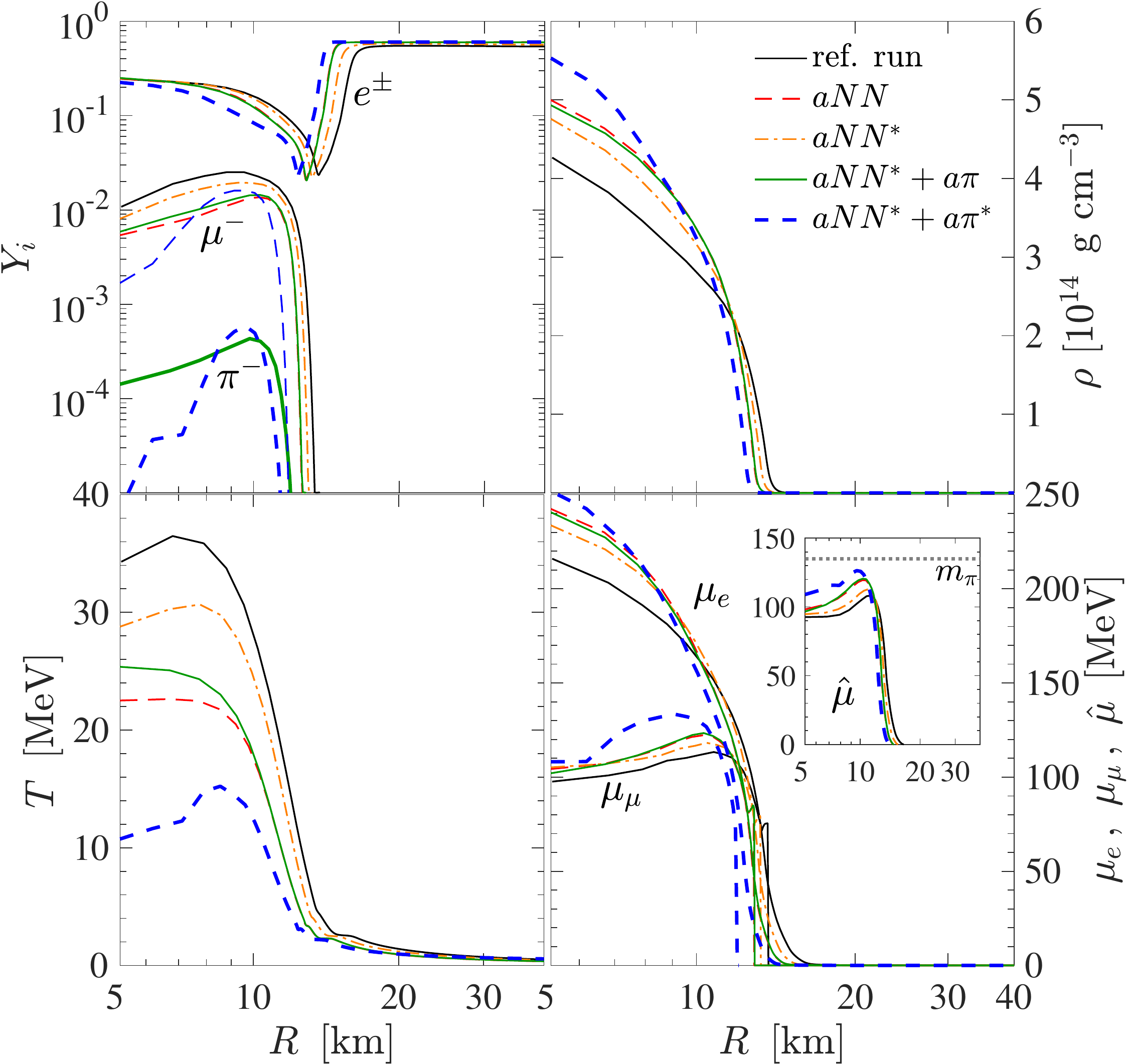}
\caption{The same as Fig.~\ref{fig:hydro_a} but at about 7~s postbounce.}
\label{fig:hydro_d}
\end{figure*}
%
\section{PNS deleptonization with axions}
\label{sec:PNS_delept_axions}
A comparison of the PNS deleptonization with muons alone and in addition with pions can be found in the Appendix~\ref{sec:PNS_delept_ref}, with an overall negligible impact on the PNS structure and evolution of the neutrino luminosities and average energies. Here, we report about simulations of the PNS deleptonization including axions. We distinguish five different SN simulation setups: (i) the reference model without axions, (ii) including axion emission from {\em only} nucleon-nucleon bremsstrahlung at the vacuum one-pion exchange level employing the axion rates of Ref.~\cite{Fischer:2016b}, denoted as $aNN$ and including all the improvements of Ref.~\cite{Carenza:2019pxu}, denoted as $aNN^*$, then further (iii) $aNN^*$ plus axions stemming from noninteracting and (iv) interacting pions, henceforth denoted as $a\pi$ and $a\pi^*$, respectively.  Table~\ref{tab:SN_setup} summarizes these different setups, including the labels used throughout the manuscript and the references for the different treatments concerning the calculations of the axion emissivity. Note that all simulations include muons and the associated muonic weak interactions, according to Ref.~\cite{Fischer:2020d}.

Furthermore, having evolved our SN simulations consistently through all SN phases, we were able to assess that the axion losses do not have a significant impact prior to the SN explosion onset. Though axions begin to be produced already slightly after core bounce from both channels~\eqref{eq:NNa} and \eqref{eq:pia}, mainly because the material is shock heated as the bounce shock propagates outward reaching temperatures on the order of 10~MeV, the axion luminosities remain negligible, on the order of at most $1$--$2\times 10^{50}$~erg~s$^{-1}$ for the bremsstrahlung (depending only slightly on the treatment of the rates $aNN$ or $aNN^*$) and $1$--$10\times 10^{49}$~erg~s$^{-1}$ for axions from pions, and are, thus, significantly below those of the neutrinos. During the postbounce evolution, the continuous increase of the central density and temperature, due to the mass accretion onto the bounce shock from the still infalling material of the stellar progenitor, leads to the continuous rise of the axion luminosity. However, the axion luminosities never reach values compatible to those of the neutrinos and remain below $10^{51}$~erg~s$^{-1}$ for both bremsstrahlung and pion rates. 
Furthermore, the axion losses are dominated by bremsstrahlung processes during the postbounce evolution. 
This is associated with the still generally low abundance of pions obtained during the entire postbounce evolution, similar to those of muons. The finding that axion losses are negligible during the SN postbounce evolution prior to the SN explosion onset is consistent with previous studies (cf. Ref.~\cite{Fischer:2016b} and references therein). 

The situation changes after the SN explosion onset, i.e., when the SN shock is accelerating to increasingly larger radii and when the central PNS enters the deleptonization and later Helmholtz cooling phase. 

\begin{figure*}[t!]
\centering
\includegraphics[width=1.5\columnwidth]{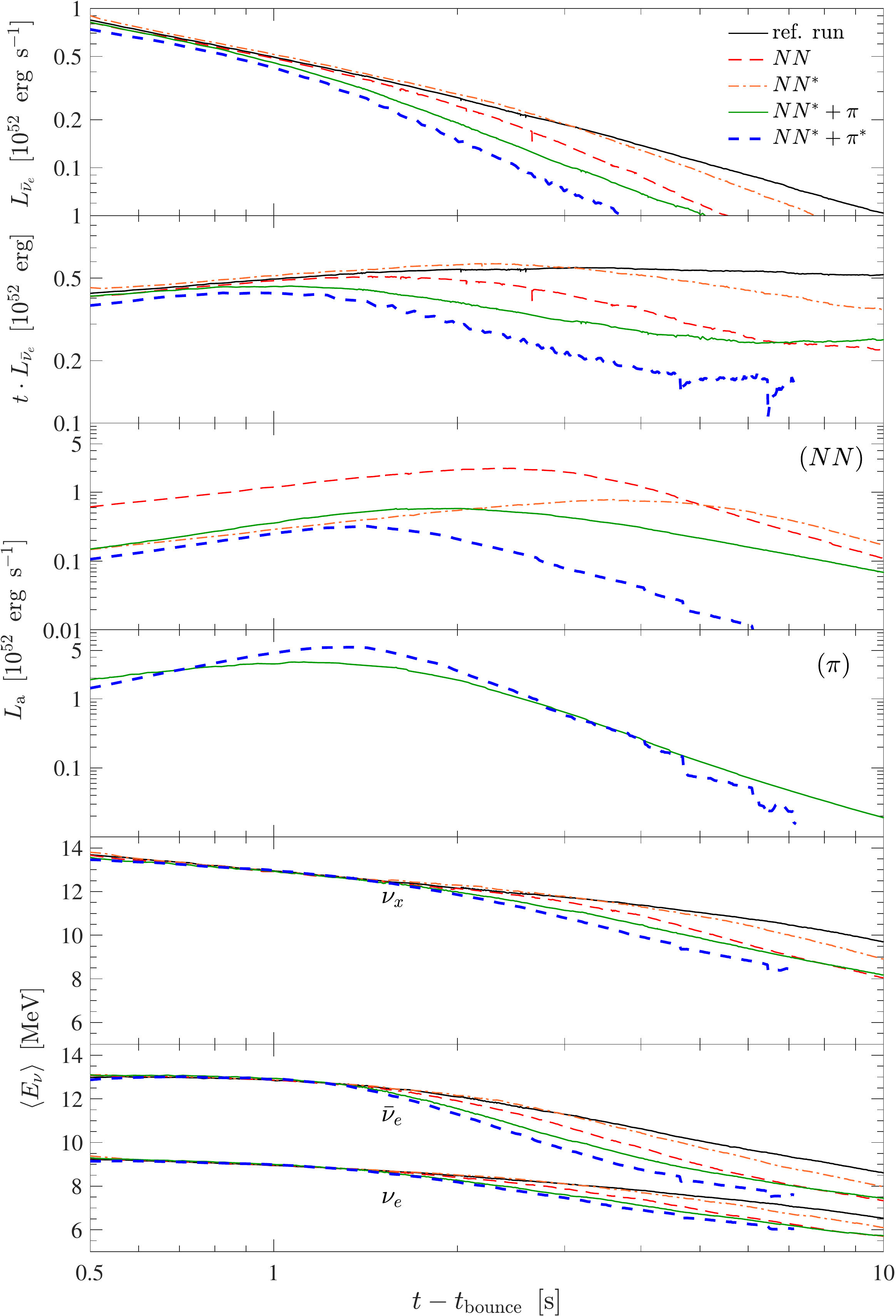}
\caption{Evolution of the $\bar\nu_e$ and axion luminosities,$L_{\bar\nu_e}$ and $L_{\rm a}$ as well as the average neutrino energies $\langle E_\nu \rangle$ for all neutrino species (here we represent the heavy neutrino flavors collectively as $\nu_x=\nu_\mu$), sampled in the comoving frame of reference at 500~km. For the axion luminosities we distinguish axions from nucleon-nucleon bremsstrahlung~\eqref{eq:NNa} ($NN$) and axions stemming from pion ($\pi$) processes~\eqref{eq:pia}.}
\label{fig:lumin}
\end{figure*}

First, we consider the role of axion losses from bremsstrahlung processes~\eqref{eq:NNa}. With the axion treatment $aNN$, at about 1~s post bounce differences are small compared to the reference simulation without axions (see Fig.~\ref{fig:hydro_a}). However, already at about 5~s post bounce the axion emission causes the PNS deleptonization evolution to accelerate in comparison to the reference case. The result is an enhanced deleptonization and, hence, compression of the PNS, featuring lower peak temperatures and a higher central density. However, the comparison of these results with the updated axion emissivity $aNN^*$ clearly states a strong overestimation of the axion losses with respect to the simplified bremsstrahlung rates $aNN$. The improved rates,$aNN^*$ leave a substantially weaker impact on the PNS deleptonization, as illustrated in Figs.~\ref{fig:hydro_a} and \ref{fig:hydro_d} at two selected times during the PNS deleptonization. Note the substantially slower deleptonization, given by a slower temperature drop, for $aNN^*$ in comparison to $aNN$.  

Consequently, the axion luminosity for $aNN$ is substantially higher than for $aNN^*$ by nearly one order of magnitude, as illustrated in Fig.~\ref{fig:lumin}. The associated accelerated PNS compression and cooling results in a shortening of the neutrino emission timescale. In turn, the neutrino luminosities and average energies of all flavors are reduced as soon as axion losses become significant. For the sake of simplicity, in Fig.~\ref{fig:lumin} we show only the neutrino luminosity for $\bar\nu_e$, as those are the most relevant ones for possible detection prospects through the inverse $\beta$ decay, e.g., at the Super-Kamiokande detector~\cite{Wu:2015}. The magnitude of the reduced neutrino emission timescale depends on the magnitude of the axion emissivity; high rates $aNN$ result in a more sever shortening which is substantially less pronounced for the updated and reduced rates $aNN^*$. Note that the latter neutrino fluxes and average energies are nearly indistinguishable from the reference case without any axion losses, during the early PNS evolution. 

Note further that since the nucleon-nucleon bremsstrahlung rates have a strong temperature dependence, their impact starts to decrease as the PNS interior cools (here, after about 5--10~s), the axion luminosity from bremsstrahlung drops and its impact on the enhanced cooling becomes negligible. After that, the PNS evolution is entirely governed by neutrino losses. 

The inclusion of axion losses stemming from pions accelerates further the cooling of the PNS interior. Since shortly after the SN explosion onset the abundance of pions increases rapidly, the associated axion luminosity rises too. This is illustrated in Figs.~\ref{fig:hydro_a} and \ref{fig:lumin} for the non-interacting pions (solid green lines), for which we set the pion self-energy to zero. Since the axion luminosity from bremsstrahlung processes remains similar as for the simulation without pions included, also the impact on the PNS structure remains comparable (c.f. the central density at 7~s post bounce in Fig.~\ref{fig:hydro_d}). Only the central temperature is somewhat lower, indicating the additional impact from axion losses at the PNS interior from pions. Consequently, also the impact on the neutrino luminosities and average energies is comparable to the setup without axion losses stemming from non-interacting pions (see Fig.~\ref{fig:lumin}). The situation changes substantially when considering interacting pions (blue dashed lines in Figs.~\ref{fig:hydro_a} and \ref{fig:hydro_d}). The significantly higher abundance of $\pi^-$ results in an enhanced axion luminosity, which, in turn, leads to an accelerated PNS cooling. This is illustrated via the significantly lower temperatures at the PNS interior already at about 7~s in Fig.~\ref{fig:hydro_d}. Correspondingly, the faster PNS compression leads to higher central densities. However, the faster temperature drop feeds back to a faster drop of the pion abundance, which still exhibits a strong temperature dependence. Hence, also the axion luminosities drop faster as illustrated in Fig.~\ref{fig:lumin} (blue dashed lines), and the impact from axion losses on the neutrino emission terminates earlier than for the case with non-interacting pions (solid green lines). Nevertheless, the lower neutrino luminosities obtained together with the reduced average neutrino energies, in particular during the early PNS deleptonization phase, at about 2--5~s, may leave interesting prospects for the neutrino detection.  In particular we find that the luminosity power law, $L\propto t^{-1}\exp\{-(t/\tau)^\alpha\}$~\cite{Roberts:2021} with characteristic cooling timescale $\tau$ and parameter $\alpha\simeq2.5$, changes substantially. This is illustrated when considering the quantity $t\cdot L_\nu$ in Fig.~\ref{fig:lumin}, which indicates that when considering additional losses, the neutrino losses cannot be considered constant. The latter is the case only for the standard reference setup, during the early PNS deleptonization phase up to about 10~s postbounce.

\section{Observable signatures}
\label{sec:detection}
In this section we discuss the possible observable signatures of axion emission in SN, especially focusing on the case of pionic processes. A first study considering only bremsstrahlung processes was presented in Ref.~\cite{Payez:2014xsa}, which we follow closely. In the following, we will show how the modification of the SN neutrino signal, due to the emission of axions, would affect the observable neutrino signal in large underground detectors. For definiteness we focus on the future-generation Hyper-Kamiokande water-Cherenkov detector, planned in Japan with a fiducial mass of 260~kton~\cite{Hyper-Kamiokande:2018ofw}.  When needed, we will also show, for comparison, results from the current Super-Kamiokande neutrino detector, for which we assume a mass of 32~kton.

The neutrino event rate, $N_e$, at Earth can be expressed symbolically as follows~\cite{Fogli:2004ff}:
\begin{equation}
N_e = F_\nu \otimes \sigma_e \otimes R_e \otimes \varepsilon~,
\end{equation}
where the neutrino flux at Earth is convoluted with the interaction cross section $\sigma_e$ in the detector for the  production of an electron or a positron, as well as the energy-resolution function $R_e$ of the detector and the detection efficiency $\varepsilon$. The threshold of the experiment is $E_{\rm th}=5$~MeV. For simplicity, in the following, we will neglect the effect of energy-dependent features. Moreover, we assume $\varepsilon=1$ above the threshold. 

Concerning the detected neutrino fluxes, notably they are affected by peculiar flavor conversion effects. Their characterization is far from being settled due to peculiar effects in the SN interior, associated with neutrino-neutrino interactions and with matter turbulence (for a review about SN neutrino oscillation effects, see Ref.~\cite{Mirizzi:2016}). Furthermore, we are mostly interested in the PNS deleptonization phase ($t\gtrsim 1$~s) after which the spectra of all neutrino flavors become increasingly similar~\cite{Fischer:2012a}, which, in turn, reduced the impact of flavor conversions on the observable signal. Therefore, for simplicity, we will neglect also these effects, assuming unoscillated neutrino spectra at Earth. 

Water-Cherenkov detectors are mostly sensitive to SN electron antineutrinos through the inverse-beta decay process ${\bar\nu_e}+p \to n+ e^+$, which we characterize from Ref.~\cite{Strumia:2003zx}. In the following, we assume a fiducial SN at a distance $d=10$~kpc.

\begin{figure}[t!]
\vspace{0.cm}
\includegraphics[width=\columnwidth]{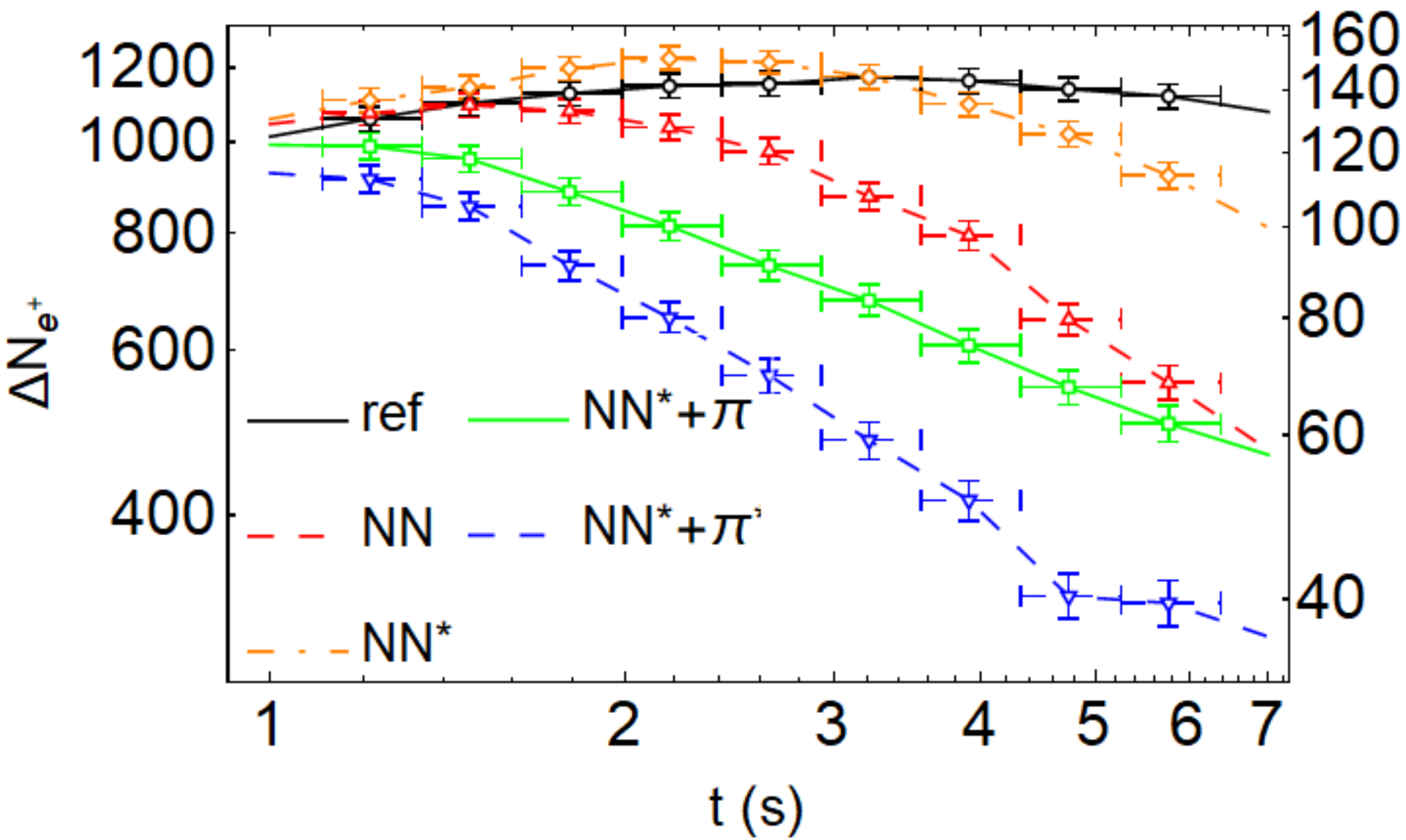}
\caption{Count rate in each log-time bin for $\bar\nu_e$ in the Hyper-Kamiokande (left scale) and Super-Kamiokande detectors (right scale)  due to the inverse-beta interaction for the different scenarios discussed in Sec.~\ref{sec:PNS_delept_axions}. The vertical error bars show the Poisson uncertainties on the counts in Hyper-Kamiokande and the horizontal error bars show the bin widths.}
\label{fig:dNdt}
\end{figure}

In Fig.~\ref{fig:dNdt}, we show the $\bar\nu_e$ event rate in the Hyper-Kamiokande detector (left scale) due to the inverse-beta interaction for the different scenarios discussed in sec.~\ref{sec:PNS_delept_axions} (see Fig.~\ref{fig:lumin}). For comparison, the right scale refers to the event rate in the Super-Kamiokande detector. In particular, following Ref.~\cite{Roberts:2021}, we adopt a log scale for the time axis, choosing ten equal-width bins per factor of 10 in time. Indeed, using a logarithmic time axis, one can use the following algebraic relation: $dA/dt= dA/d\,{\rm ln} t= (2.3)^{-1} dA/d\,{\rm log}_{10} t$.  Therefore, Fig.~\ref{fig:dNdt} shows $\Delta N=(2.3)^{-1} dN/d\,{\rm log}_{10} t$. We see that in the reference run without axion emission (black line) the behavior of $\Delta N$ is rather constant with time, for the interval considered here. Including axion emission from bremsstrahlung process in the $aNN$ treatment would cause a fast drop already at $t \gtrsim 2$~s, leading to a reduction by a factor of about 2 of the signal at $t\sim 7$~s.  The updated bremsstrahlung rate $aNN^*$ has a considerably reduced impact and the complete run becomes nearly indistinguishable from the reference one. Differences become visible only at $t\gtrsim 4$~s, reducing the signal by a factor of about $1.2$ with respect to the reference case. If we now include the pionic process, the drop of the neutrino event rate becomes dramatic, being evident with respect to the reference case already at about $t\gtrsim 1$~s. In the case of non-interacting pions $a\pi$ the final reduction of the signal is comparable to what is obtained with the $aNN$ bremsstrahlung process. However, the drop of the signal is faster at early times. Including interacting pions $a\pi^*$ features an even more dramatic drop of the event rate at early times, resulting in a reduction by a factor of 3 at $t\sim 7$~s with respect to the reference case.

\begin{figure}[t!]
\vspace{0.cm}
\includegraphics[width=\columnwidth]{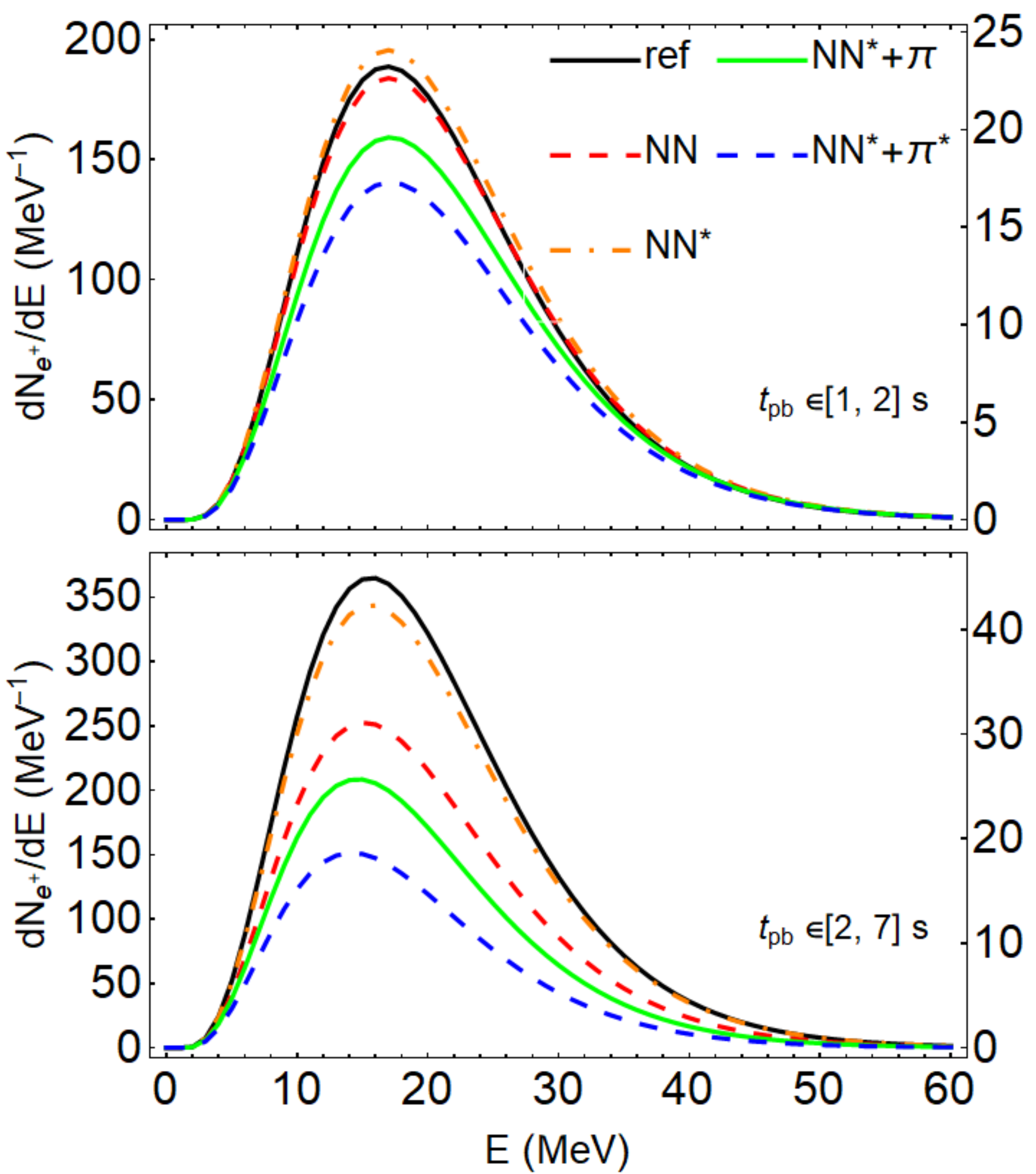}
\caption{Positron spectra in the Hyper-Kamiokande (left scale) and Super-Kamiokande detectors (right scale) in selected time ranges, between $1$--$2$~s (upper panel) and $2$--$7$~s (bottom panel) for the models discussed in Sec.~\ref{sec:PNS_delept_axions}.}
\label{fig:dNdE}
\end{figure}

Figure~\ref{fig:dNdE} shows the detectable positron spectra for the different models considered, integrating the signal in two time windows, namely $t\in[1;2]~s$ (upper panel) and $t\in[2;7]~s$ (lower panel), respectively. We see that at early times the effect of axion emission is marginal in the case of only bremsstrahlung processes, regardless of whether the rate is corrected ($aNN^*$) or not ($aNN$). On the other hand, in the presence of pionic processes, the suppression of the positron spectrum already shows a reduction of a factor about $1.6$ at the peak with respect to the reference case, for the interacting pion scenario $a\pi^*$. In the second, later time window the effect of the suppression of the positron spectrum is more remarkable, except in the case of $aNN^*$, which remains nearly indistinguishable in comparison to the reference case. On the other hand, we find a reduction of the peak of the spectrum of a factor of about 3 in the case of interacting pions $a\pi^*$.

\begin{figure}[t!]
\vspace{0.cm}
\includegraphics[width=\columnwidth]{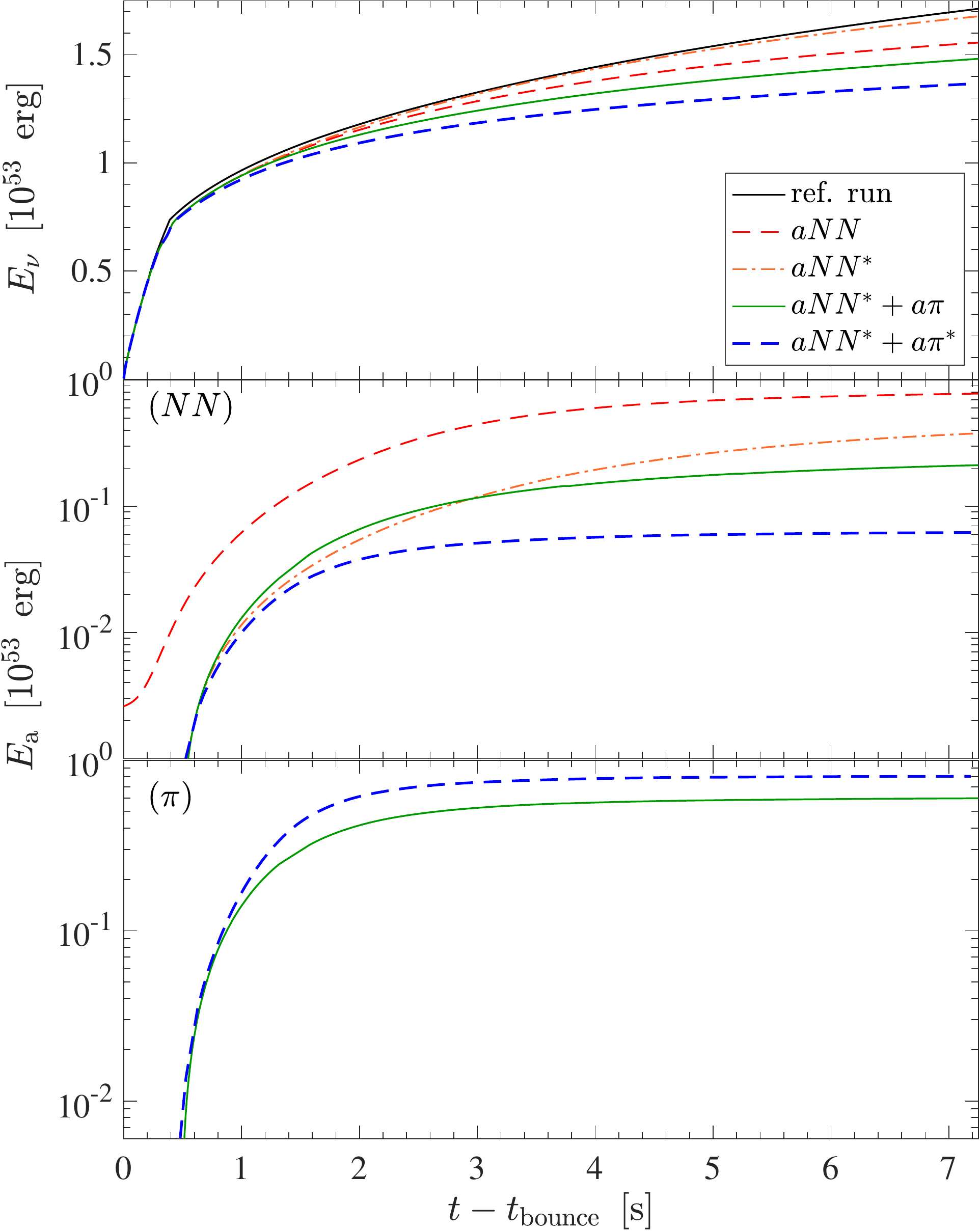}
\caption{Evolution of the SN binding energy carried by neutrinos (upper panel) and axions due to nucleon-nucleon bremsstrahlungs processes (middle panel) and stemming form pions (lower panel), for the scenarios discussed in Sec.~\ref{sec:PNS_delept_axions}.}
\label{fig:ebind}
\end{figure}

A further signature of the axion emission would be the impact on the SN binding energy, $E_B$, carried by neutrinos. Note that the axion emission would lead to a reduction of the SN binding energy. This effect is illustrated in Fig.~\ref{fig:ebind}, where we show the evolution of the energy carried away to infinity by neutrinos of all flavors (top panel) and axions, distinguishing bremsstrahlung processes (middle panels) and pionic processes (bottom panel), for the different scenarios discussed above. Note that the early and sudden rise of the energy carried by neutrinos is due to the contributions from the stellar core collapse, bounce, and early postbounce phases prior to the SN explosion onset, during which all simulations are identical and during which we did not find any impact from the inclusion of axions. 

We realize that in the case of bremsstrahlung processes the energy carried by axions rises continuously, as indicated by the slowly increasing temperature at the PNS center where most axions are being produced from bremsstrahlung processes. When the temperature starts to decrease, the axion emissivity decreases simultaneously, which is the case at around 5~s for $aNN$ and only at around 10~s for $aNN^*$. The total energy carried away is similar for $aNN$ and $aNN^*$; however, both of their magnitudes, $8\times 10^{52}$ and $4\times 10^{52}$~erg, respectively, are much lower than that of neutrinos, i.e., neutrinos dominate the losses during the PNS deleptonization phase. In particular, the weak feedback for $aNN^*$ leaves nearly no difference notable in the energy carried away by neutrinos, in comparison to the reference case. 

Contrarily, in the case of pionic processes a substantially larger fraction of the binding energy, comparable to the energy carried by neutrinos, is carried by axions already at early times. However, the associated rapid cooling contributions lead to a fast reduction of the PNS interior temperature, which, in turn, result in a faster drop of the axion emissivity.  Hence, the process of energy loss through axions stemming from pionic processes ceases much earlier compared to the case of from bremsstrahlung processes. Consequently, the impact on the total energy carried by neutrinos is small, a factor of about 1.3 at 7~s.  The total energy carried away by neutrinos and axions is listed in Table~\ref{tab:ebind}, evaluated at about 20~s post bounce for the different simulations. While for the axion losses the asymptotic values have been reached within the first 7~s after bounce, for the neutrino losses we extrapolate assuming monotonically decreasing neutrino luminosities of all flavors.

\begin{table}[t!]
\caption{Total energy carried away by neutrinos, $L_\nu$, and axions, $L_{\rm a}$, evaluated at a postbounce time of 20~s.}
\centering
\begin{tabular}{l c c c}
\hline
\hline
Model & $E_\nu$ & $E_{\rm a}^{NN}$ & $E_{\rm a}^{\pi}$  \\
 & $[10^{53}\,\,\,{\rm erg}]$ & $[10^{53}\,\,\,{\rm erg}]$ & $[10^{53}\,\,\,{\rm erg}]$ \\
\hline
Ref. & 2.2 & \ldots & \ldots \\
$aNN$ & 1.8 & 0.88 & \ldots \\
$aNN^*$ & 2.0 & 0.58 & \ldots \\
$aNN^*+a\pi$ & 1.7 & 0.29 & 0.79 \\
$aNN^*+a\pi^*$ & 1.5$^\dagger$ & 0.06 & 0.81 \\
\hline
\end{tabular}
\label{tab:ebind}
\\
$^\dagger$: Asymptotic values extrapolated assuming monotonically decreasing neutrino luminosities for all flavors beyond 7~s postbounce evolution.
\end{table}

\begin{figure}[t!]
\vspace{0.cm}
\includegraphics[width=\columnwidth]{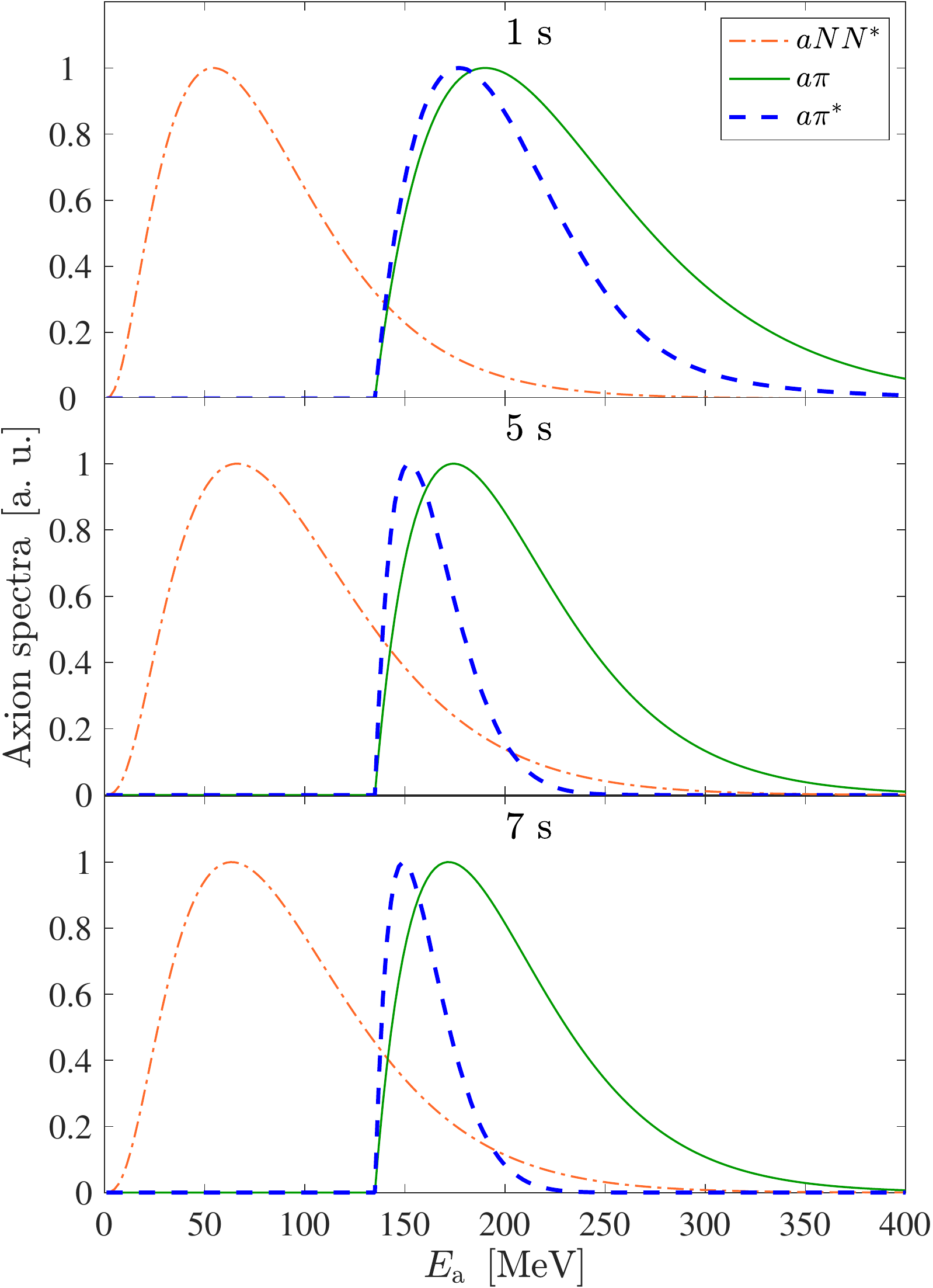}
\caption{Normalized axion-emission spectra for nucleon-nucleon bremsstrahlung processes (light red dash-dotted lines) and for axions stemming from noninteracting pions (solid green lines) and interacting pions (thick blue dashed lines), at different postbounce times 1 (top panel), at 5 (middle panel) and at 7~s (bottom panel), corresponding to the SN simulations discussed in Sec.~\ref{sec:PNS_delept_axions}.}
\label{fig:pionspectr}
\end{figure}

What is described above are all indirect signatures of the axion effects in a SN. However, a direct detection of SN axions is perhaps also possible. To explore this intriguing possibility, in Fig.~\ref{fig:pionspectr} we show the axion spectrum, integrated over the entire PNS profile, for the bremsstrahlung and pionic processes, distinguishing non-interacting pions and strongly interacting pions, evaluated at different postbounce times. Two features are clearly visible in the pion-induced axion spectrum: (i) the energy threshold at the pion mass and (ii) the energy peak at $\sim$150--200~MeV. Both features can be understood from the kinematics of the process \eqref{eq:pia}. Assuming equal mass of protons and neutrons, it is evident that the minimal axion energy corresponds to the minimal pion energy and, hence, with its mass. This is true also when the pion interactions are turned on, and the pion dispersion relation is corrected with $\Sigma_{\pi^-}$ [cf. Eq.~\eqref{eq:pion-selfenergy} and Fig.~\ref{fig:sigma}]. The higher threshold also moves the peak energy to higher values with respect to the bremsstrahlung process.

Notice also the modified dispersion relation \eqref{eq:pion-selfenergy}, with $ \Sigma_{\pi^-} $ becoming increasingly more negative at large momenta, which, in turn, makes it more expensive to increase the pion energy. Since the axion energy in the reaction \eqref{eq:pia} is controlled by the pion energy (the nucleons are much heavier), this reflects in a narrower axion spectrum when the pion interactions are turned on ($\Sigma_{\pi^-}\neq 0$). This feature is clearly evident in the spectra in Fig.~\ref{fig:pionspectr}  (cf. the green line, which shows the axion spectrum for non-interacting pions, with the blue dashed line, for which the interactions are turned on).

It is also apparent that at early times, when the PNS is still hot and rather dilute with large radii, the axion spectra can have extended tails up to 1~GeV. It was estimated in Ref.~\cite{Carenza:2020} that for a close-by SN, i.e., $d \lesssim 1$~kpc, and a megaton-class water-Cherenkov underground detector, this axion burst would be detectable through the processes $a+N \to N + \pi$, giving an unexpected signal of ${\mathcal O}(10^3)$ pions during a SN emission. The detection of this pionic signal, together with a rapid drop of the neutrino signal, would be strong evidence for the axion emission via pionic processes.

\section{Summary and conclusions}
\label{sec:summary}
Results are reported from simulations of the PNS deleptonization phase of core-collapse SN, launched from a progenitor of 18~$M_\odot$~\cite{Woosley:2002zz}, where in addition to the neutrinos, axions and the associated energy losses are considered. The SN simulations are based on general relativistic neutrino radiation hydrodynamics in spherical symmetry, employing six-species Boltzmann neutrino transport. The restriction to spherical symmetry is justified for the PNS deleptonization phase of a core-collapse SN, since  the compact central remnant PNS obeys spherical symmetry. The description of approximate convection can be implemented in such spherically symmetric models~\cite{Roberts:2012c}. All present simulations include muons as an additional degree of freedom, and the associated muonic weak processes~\cite{Guo:2020,Fischer:2020d}.

In a first step, updated nucleon-nucleon bremsstrahlung rates for the axion emissivity are considered~\cite{Carenza:2019pxu}. These yield a substantial reduction of the previously reported shortening of the neutrino emission timescale~\cite{Fischer:2016b}, which adopted the simplified bremsstrahlung vacuum rate expressions based on the one-pion exchange approximation. In a second step, the inclusion of pions enables us to extend the previous studies including axions stemming from pions. This novel channel has long been omitted, due to the shortcoming of a generally negligible pion abundance. Considering interacting pions, however, relaxes this argument. Then, the inclusion of axion cooling from this channel significantly shortens the timescale of neutrino emission as a direct feedback, in comparison to the case when only bremsstrahlung processes are considered. 

The resulting SN simulation neutrino signals are then further analyzed with respect to the detection prospects at the future-planned Hyper-Kamiokande and presently operating Super-Kamiokande water-Cherenkov detectors. We find that the sudden drop of the count rate implies the presence of additional, nonstandard energy losses, that could be related with axions originating from pions due to Compton processes, if occurring during the early phase of detection.

Furthermore, according to the recent study of Ref.~\cite{Reed:2020gva}, a measurement of the binding energy of less than $1.5 \times 10^{53}$ erg carried by neutrinos, would be strong evidence for the presence of exotic energy losses. It would be otherwise incompatible with SN models for different neutron star mass and EOS. An experiment like Hyper-Kamiokande has the potential to measure the total neutrino energy with an accuracy of few $\%$~\cite{GalloRosso:2017mdz}, allowing us to diagnose with accuracy the presence of an additional energy-loss.

Note that the uncertainty due to the elastic approximation, i.e., zero nucleon recoil, implemented for the calculation of the emissivity of axions stemming from pions, is unlikely to alter the findings of the present work. Even though this gives rise to a sharp cut of the rate given by the pion rest mass, the high-energy tail of the axion spectrum dominates the phase space integrals and hence the axion emissivity. 

It is important to note that all these results, as well as those reported in Ref.~\cite{Fischer:2020d}, with a negligible impact of muons and associated muonic weak processes during the SN bounce and early postbounce evolution, are based on the HS(DD2) relativistic mean-field EOS. It represents a stiff nuclear model at densities in excess of saturation density. 
These findings, e.g., the negligible impact from the inclusion of muons and pions reported here, may alter when adopting a softer nuclear EOS, i.e. when higher central densities are encountered~(see Refs.~\cite{Bollig:2017,Bollig:2020,Steiner:2013} and the left panel of Fig.~\ref{fig:central_appendix} in the Appendix~\ref{sec:PNS_delept_ref}). This yet incompletely understood aspect remains to be explored in a more systematic fashion. Besides the uncertainty of the EOS in terms of bulk properties, there are other aspects that are likely to affect the signal proposed here, e.g., impact of the EOS on the neutrino opacity, PNS convection, and fall back. All this remains to be explored in a systematic fashion.

We note further that magnetic fields are not considered here, due to the restriction to spherical symmetry. The magnetic field can be large at the interior of PNS~\cite{Braithwaite:2004,Takiwaki:2009}, which is usually associated with magnetorotationally driven SN explosions~\cite{LeBlanc:1970kg,Winteler:2012}, giving rise to the formation of magnetars---an observationally known class of highly magnetized neutron stars with surface magnetic field on the order of $10^{15}$~G. This is still an active subject of research, in particular regarding the development of jets in the context of core-collapse SNe~\cite{Moesta:2014,Moesta:2015}, especially the  magnitude of the initial progenitor magnetic field strength and the possible amplification during the SN evolution. Furthermore, it is expected that strong magnetic fields affect the emission of axions~\cite{Kachelriess:1997,Maruyama:2018}. An interesting possibility is the possible mixing of the axions with photons~\cite{RaffeltStodolsky:1988}, for which it has been demonstrated recently that the axion mixing with the photon longitudinal mode is suppressed for light axions due to the electron degeneracy~\cite{CaputoCarenzaLucente:2021}. The only possibility is the mixing with the photon transverse mode which requires the photon mass to match the axion mass. Hence, in the present study we consider light axions for which the mean free path is large compared to the PNS radius, even in the presence of large magnetic fields. Revisiting the impact of magnetic fields in detail would extend beyond the scope of the present study, which identifies possibly observable signatures during the PNS deleptonization, that originate from enhanced cooling contributions associated with axion emission. We postpone the full study of magnetic field impact on SN axions to a future work, which must include self-consistently the effect of magnetic fields not only on the axion emission but also on the PNS structure and the associated neutrino emission. 

\begin{acknowledgments}
T.F. acknowledges support from the Polish National Science Center (NCN) under Grant No.~2020/37/B/ST9/00691.  The work of P.C. and A.M. is partially supported by the Italian Istituto Nazionale di Fisica Nucleare (INFN) through the ``Theoretical Astroparticle Physics'' project and Research Grant No.~2017W4HA7S ``NAT-NET: Neutrino and Astroparticle Theory Network'' under the program PRIN 2017 funded by the Italian Ministero dell'Universit\`a e della Ricerca (MUR). B.F. acknowledges support from the SciDAC Grant No.~A18-0354-S002 (de-sc0018232). The work of S.R. was supported by the U.S. Department of Energy under Grant No.~DE-FG02-00ER41132. The work of M.G. was supported by funding from a grant provided by the Fulbright U.S. Scholar Program. M.G. thanks the Departamento de Física Teórica and the Centro de Astropartículas y Física de Altas Energías (CAPA) of the Universidad de Zaragoza for hospitality during the completion of this work.  This work was supported by the COST Actions No. CA16117 “ChETEC” and No. CA16214 “PHAROS”. The SN simulations were performed at the Wroclaw Center for Scientific Computing and Networking (WCSS) in Wroclaw (Poland).
\end{acknowledgments}

\begin{figure*}[t!]
\centering
\subfigure[]{\includegraphics[width=\columnwidth]{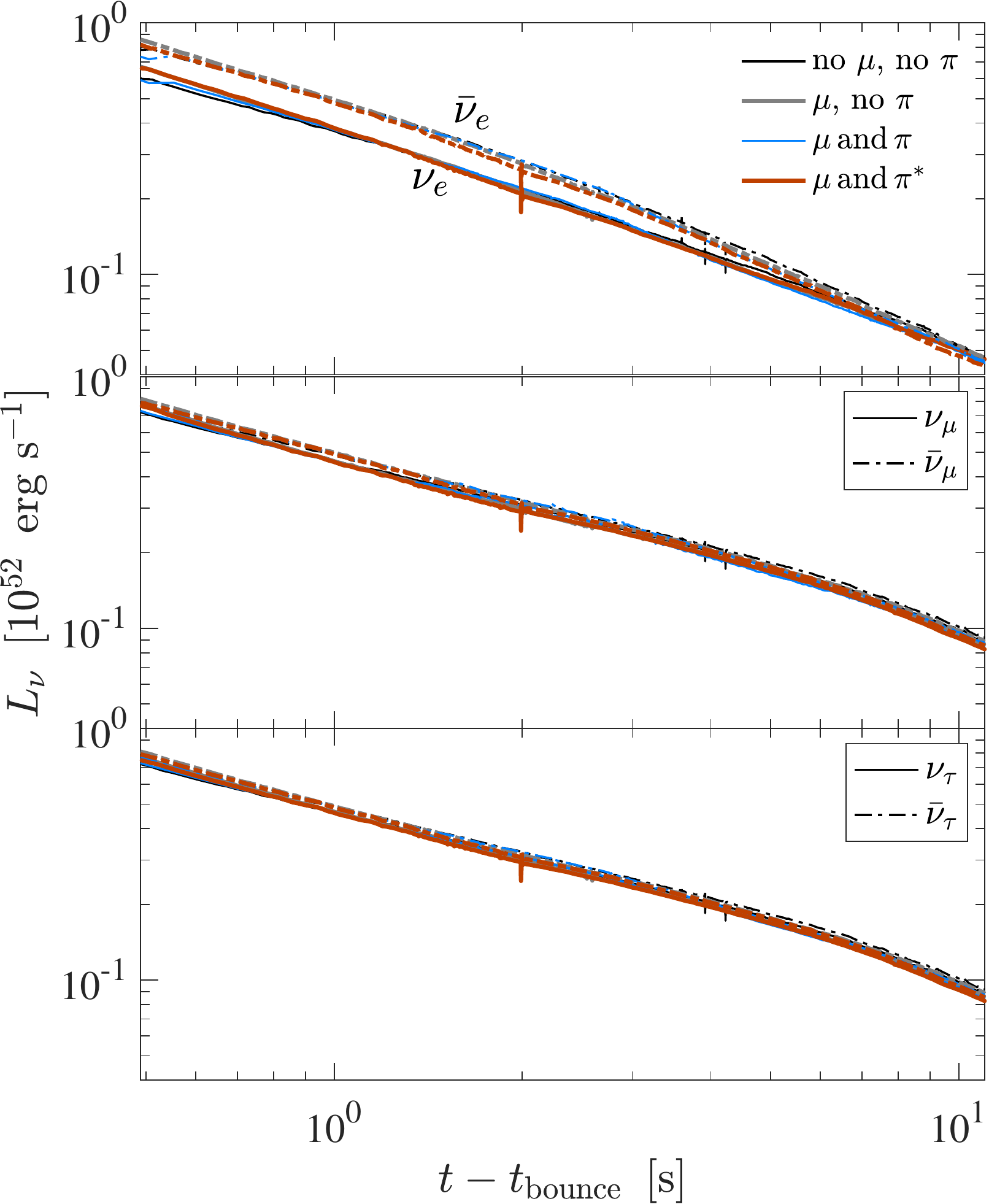}\label{fig:lumin_appendix}}
\hfill
\subfigure[]{\includegraphics[width=0.96\columnwidth]{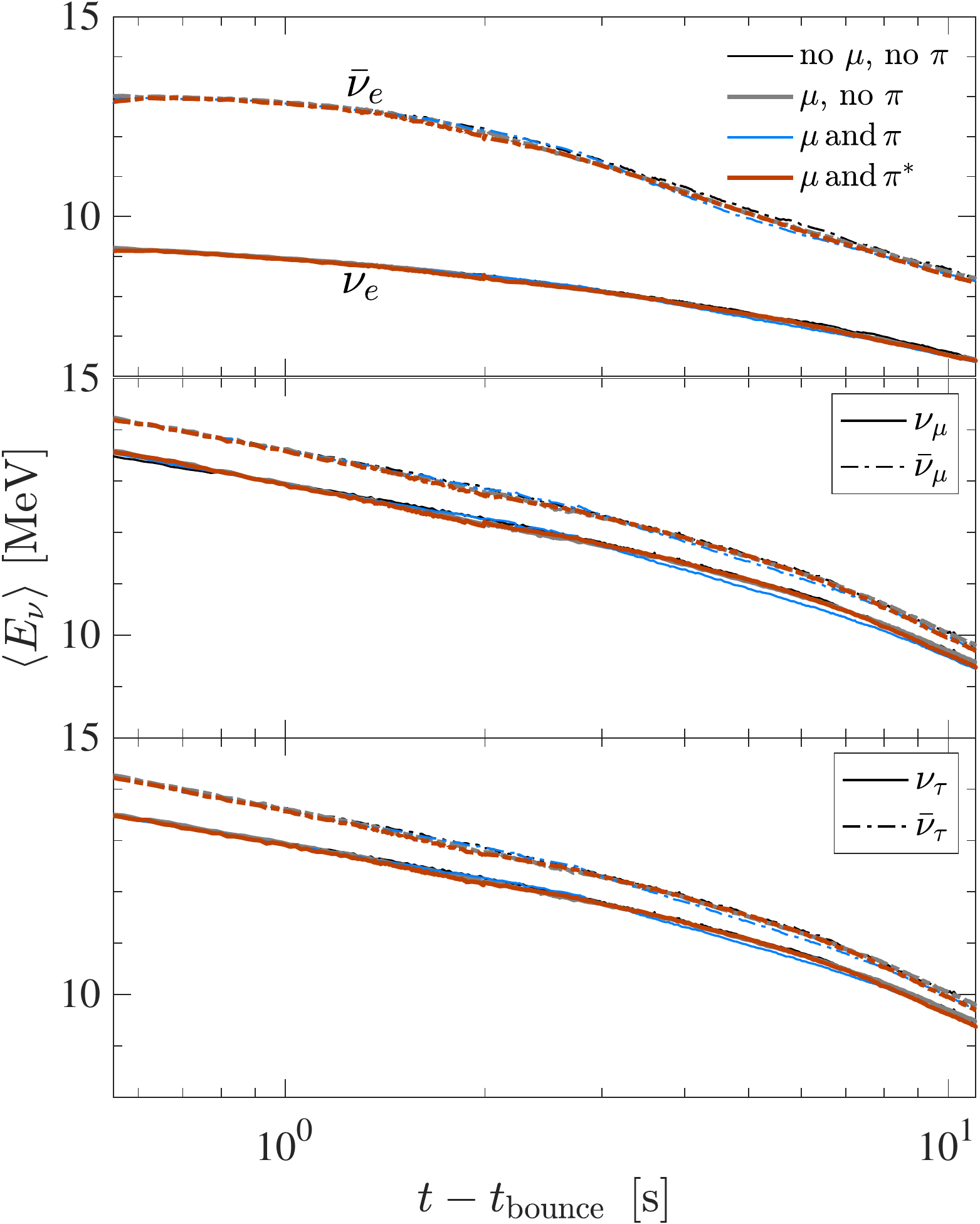}\label{fig:rms_appendix}
}
\caption{Evolution of the neutrino luminosities and average energies sampled in the comoving frame of reference at 500~km, comparing the simulations without muons and without pions (black lines) and including muons but no pions (gray lines), muons and noninteracting pions (blue lines) as well as muons and interacting pions (brown lines). All simulations employ the six-species neutrino transport scheme.}
\end{figure*}
\begin{figure*}[htp]
\includegraphics[width=2\columnwidth]{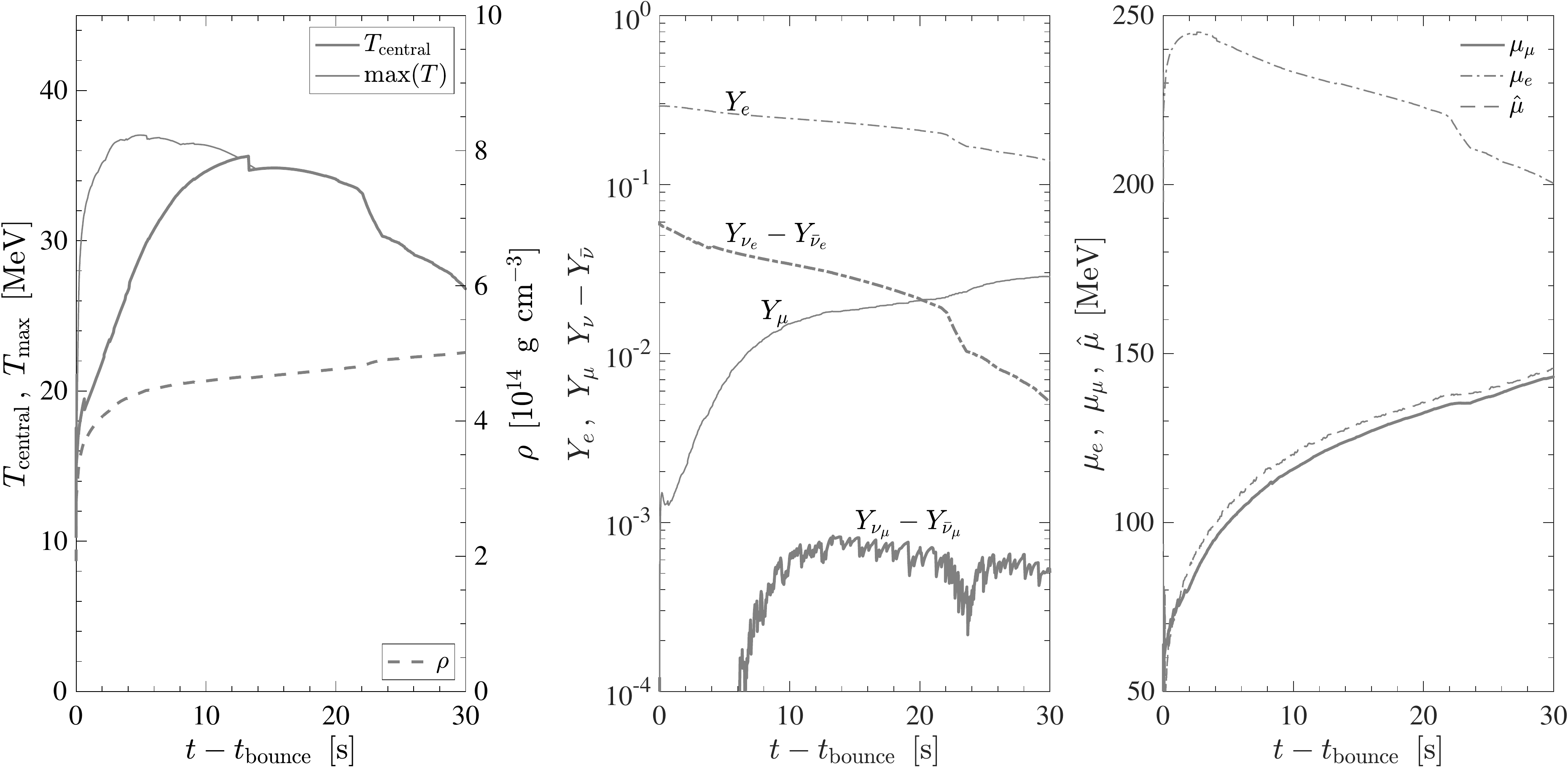}
\caption{Evolution of selected central quantities during the PNS deleptonization of the reference simulation including muons and associated muonic weak processes. Left panel:~central temperature $T_{\rm central}$ (thick solid line), in comparison to the maximum temperature ${\rm max}(T)$ (thin solid line), and central density $\rho$ (dashed line). Middle panel:~electron and muon abundances, $Y_e$ (thin dash-dotted line) and $Y_\mu$ (thin solid line), and the net neutrino abundances for $\nu_e$ (thick dash-dotted line) and $\nu_\mu$ (thick solid line). {\em Right panel:}~chemical potentials for muons, $\mu_\mu$ (thick solid line),  electrons, $\mu_e$ (thin dash-dotted line), and $\hat\mu$ (thin dashed line).}
\label{fig:central_appendix}
\end{figure*}
%

\appendix
\section{PNS DELEPTONIZATION WITH MUONS AND PIONS}
\label{sec:PNS_delept_ref} 
A comparison is presented here for the simulations of the PNS deleptonization phase, comparing the following four different setups: (i) without muons and without pions, henceforth denoted as (no $\mu$, no $\pi$), (ii) including muons and associated muonic weak processes but without pions ($\mu$, no $\pi$), (iii) with muons and noninteracting pions (no $\mu$ and $\pi$), and (iv) with muons and interacting pions following Ref.~\cite{Fore:2020} ($\mu$ and $\pi^*$). The discussion about the implementation of the pions is given in Sec.~\ref{sec:SNmodel}, concerning both treatment for interacting and noninteracting pions. All simulations are based on the six-species Boltzmann neutrino transport scheme of {\tt AGILE-BOLTZTRAN}~\cite{Fischer:2020d}. The standard set of weak reactions used here is listed in Table~I of Ref.~\cite{Fischer:2020a}, including the corresponding references. Furthermore, we use the fully inelastic charged-current electronic and muonic rates of Ref.~\cite{Guo:2020}. The latter include self-consistent contributions of weak magnetism.

The SN simulations are launched from the 18~$M_\odot$ progenitor from the stellar evolution series of Ref.~\cite{Woosley:2002zz}. It has been evolved through all SN phases self-consistently. Since the present focus is on the PNS deleptonization, and since neutrino-driven SN explosions cannot be obtained in spherical symmetry for this class of iron-core progenitors, we enhance the charged-current electronic weak reaction rates in the gain layer, applying the procedure introduced in Refs.~\cite{Fischer:2009af,Fischer:2012a}. It results in an explosion onset at about 350~ms postbounce, defined when the SN bounce shock reaches a radius of about 1000~km. Note that at that stage the enhancement of the charged-current weak rates is turned off and the standard weak rates setup is used again. The remnant PNS contain about 1.54~$M_\odot$ of baryonic mass.

Figure~\ref{fig:lumin_appendix} shows the resulting evolution of the neutrino luminosities, and Fig.~\ref{fig:rms_appendix} shows the average energies, sampled in the comoving frame of reference at a radius of 500~km. The simulations of the early PNS deleptonization phase, up to about 10~s postbounce, which include muons and associated muonic weak reactions~\cite{Guo:2020,Fischer:2020d}, show a negligible deviation from the reference setup without muons. This is attributed to two facts: (i)~The early PNS deleptonization phase is governed by neutrino losses at the PNS surface, where the neutrinospheres for all flavors are located, still at moderately low densities on the order of $<10^{13}$~g~cm${-3}$, while muons are abundant at higher densities, and (ii)~the abundance of muons is generally too low to leave a notable impact on the softening of the high-density EOS. We report that only at later times, $>10$~s postbounce, do muons start to impact the PNS deleptonization and cooling, by means of an accelerated PNS compression. 

A second set of simulations focuses on the role of pions. The latter are in thermal and chemical equilibrium. Note that only $\pi^-$ are considered; $\pi^+$ and $\pi^0$ are strongly suppressed. Here we consider noninteracting pions, which obey the standard thermal bosonic distribution, and interacting pions following Ref.~\cite{Fore:2020}. The latter gives rise to a nontrivial pion dispersion relation featuring the pion self-energy. The inclusion of noninteracting pions leaves a negligible impact on the early PNS deleptonization phase, for the same reasons (i) and  (ii) for muons aforementioned and, moreover, because the abundance of noninteracting pions is lower, by more than one order of magnitude, than the abundance of muons. The situation changes only marginally when considering interacting pions, whose abundance reaches, and even exceeds, those of the muons at the center of the PNS. However, their overall impact on the PNS structure and evolution during the early (about 1--10~s) PNS deleptonization remains negligible. An impact on the neutrino fluxes and spectra could not be found. 

It is interesting to note that toward later times during the PNS deleptonization, on the order of 20--30~s, when the central temperature starts to decrease the simulation enters the cooling phase (see the left panel in Fig.~\ref{fig:central_appendix}). Relatedly, the central net neutrino abundance of the electron flavors starts to decrease. Furthermore, note that only after that, do the chemical potentials of muons approach the one of the electrons (see the right panel in Fig.~\ref{fig:central_appendix}); i.e., weak equilibrium will be established only at post bounce times greater than 30~s when $\mu_\mu\simeq\mu_e$.

\bibliographystyle{utphys}
\providecommand{\href}[2]{#2}\begingroup\raggedright\endgroup

\end{document}